\documentclass[reprint,superscriptaddress, amsmath,amssymb, floatfix, pra]{revtex4-2}

\usepackage{xcolor}
\usepackage{graphicx}
\usepackage{braket}
\usepackage{mathtools}
\usepackage[colorlinks, citecolor=blue]{hyperref}
\usepackage[utf8]{inputenc}
\usepackage[T1, OT1]{fontenc}
\usepackage{dsfont}

\begin{document}
\title{Optical realization of one-dimensional generalized split-step quantum walks}
\author{P. A. Ameen Yasir}
\email{ameenyasir@iisc.ac.in}
\author{Abhaya S. Hegde}
\email{abhayhegde16@alumni.iisertvm.ac.in}
\affiliation{Quantum Optics \& Quantum Information,  Department of Instrumentation and Applied Physics, Indian Institute of Science, Bengaluru 560012, India}
\author{C. M. Chandrashekar}
\email{chandracm@iisc.ac.in}
\affiliation{Quantum Optics \& Quantum Information,  Department of Instrumentation and Applied Physics, Indian Institute of Science, Bengaluru 560012, India}
\affiliation{The Institute of Mathematical Sciences, C. I. T. Campus, Taramani, Chennai 600113, India}
\affiliation{Homi Bhabha National Institute, Training School Complex, Anushakti Nagar, Mumbai 400094, India}

\date{\today}

\begin{abstract}
Quantum walks are more than tools for building quantum algorithms. They have been used effectively to model and simulate quantum dynamics in many complex physical processes. Particularly, a variant of discrete-time quantum walk known as split-step quantum walk is closely related to Dirac cellular automata and topological insulators whose realizations rely on position-dependent control of evolution operators. Owing to the ease of manipulating multiple degrees of freedom of photons, we provide an optical setup of split-step operators which in combination with position-dependent coin (PDC) operation can accomplish a table-top setup of generalized split-step walks. Also, we propose an optical implementation for PDC operation that allows, for instance, to realize electric quantum walks, control localization dynamics, and emulate space-time curvature effects. In addition, we propose a setup to realize {\it any} $t$-step split-step quantum walk involving 2 $J$-plates, 2 variable waveplates, a half-waveplate, an optical switch, and an optical delay line.
\end{abstract}



\maketitle

\section{Introduction}
Quantum walks offer simpler means to contrive quantum simulators, allowing one to probe complex and directly inaccessible quantum processes in a controllable manner. Since quantum walks offer universal quantum computation, they go beyond just being quantum counterparts to classical random walks\,\cite{andraca2012}. 
Although known for spreading quadratically faster in position space, they have also proved to be of immense utility in various domains, like in quantum computation\,\cite{Chi09, SCS21}, quantum algorithms\,\cite{Rportugal13}, generating quantum random numbers\,\cite{SC19}, modeling topological phenomena\,\cite{kitagawa2010, COB15}, simulating Floquet-Chern insulator\,\cite{errico2020}, simulation of Majarona modes and edge states\,\cite{zhang2017}, studying interplay between Bloch oscillations and Landau-Zener transitions\,\cite{errico2021}, and entanglement transfer protocols\,\cite{giordani2021}. In parallel to the various theoretical proposals highlighting the practical usefulness of quantum walks, significant progress has also been reported demonstrating the experimental implementation and controlling the dynamics of discrete-time quantum walks (DTQW) in several quantum systems, such as linear optics\,\cite{broome2010}, ion traps\,\cite{zahringer2010,alderete2020}, and neutral atom traps\,\cite{karski2009}. 

The ever increasing interest in quantum walks as theoretical models for exploring various complex phenomena naturally begets the implementation schemes suitable for a laboratory setting~\cite{KJ14}. The photonic systems have proved to be one of the most viable platforms for witnessing DTQWs at room temperature, in that they have been demonstrated under various settings such as, time-bin encoding\,\cite{BFB16}, interferometers\,\cite{AW12}. Quantum walks using spin angular momentum (SAM) and orbital angular momentum (OAM) of light beams are also proposed\,\cite{EJ05, yasir2022,zhang2010} and some are experimentally demonstrated\,\cite{cardano2015, cardano2016, cardano2017, wang2018, sephton2019, giordani2019}. The cost-effectiveness of producing single photons now prompts one to favor optical setups. Furthermore, a photonic platform enables each photon in free space to act as a walker.

It is well known that a DTQW initiated in a symmetric state involving a repeated application of the same walk unitary results in a symmetric probability distribution~\cite{NV00, CSL08}. Modifying the walk unitary at each iteration of the walk gives rise to a variety of probability distributions~\cite{Cha12, PF18}. For instance, in split-step quantum walks\,(SSQW), each walk requires two step operations and two coin operations\,\cite{kitagawa2010,kitagawa2012, barkhofen2017, nitsche2019}. Varying the rotation angle in the coin operator provides one way to tweak the walk unitary in a controllable fashion. In a lattice, altering the coin operator can be done at either each step leading to step-dependent coin operation~\cite{PF18} or each position giving rise to position-dependent coin (PDC) operation~\cite{ASS20}. PDC operation has been realized in different platforms such as polarization-path degrees of freedom\,(DoF)\,\cite{kitagawa2012,bian2015,xiao2020} and polarization-time bin DoF\,\cite{schreiber2011, barkhofen2017}.

In this work, we focus on position-dependent operation owing to its unique properties being useful in various avenues such as inducing topological phases~\cite{PF21}, quantum ratchets~\cite{CDM17}, implementing generalized measurements on a single qubit\,\cite{bian2015, kurzynski2013}, and universal quantum computation using quantum walks\,\cite{lovett2010, SCS21}. This flexibility in the PDC operation has far reaching implications to the extent of signifying curvature of the space-time lattice~\cite{AFF16, mallick2019}. The electric quantum walks featuring unique transport properties can be formulated from PDC operations~\cite{cedzich2013}. Moreover, the universal quantum computing based on single-particle DTQW relies on position-dependent operation to effectively implement the phase gates~\cite{SCS21}.

We propose a linear optical realization of PDC operator and translation operators using OAM-polarization DoF of a single photon. This will allow us to build a complete setup to realize various kinds of discrete-time quantum walks on a table-top optical bench setting. 

The paper is organized as follows. We begin with a brief introduction to various kinds of quantum walks in Sec.~\ref{sq}. A proposal for coin and step operators are given in Sec.~\ref{re} which is later extended to an optical setup for generalized split-step quantum walks. We conclude with a short discussion in Sec.~\ref{di}. 

\section{Discrete-time quantum walks and PDC operator} 
\label{sq}

The evolution of the DTQW is defined and controlled using a unitary operator defined on a tensor product of two Hilbert spaces $\mathcal{H}_c \otimes  \mathcal{H}_p$, where $\mathcal{H}_c$ is the coin Hilbert space spanned by the internal states $\vert 0\rangle$ and $\vert 1\rangle$ of a walker (a single qubit), while $\mathcal{H}_p$ represents the position Hilbert space given by the position states $\vert x \rangle$ with $x  \in \mathbb{Z}$. Here, the unitary quantum coin operation, $\hat{C}_\theta$, is a unitary rotation operator that acts only on the coin qubit space, 
\begin{align}
\hat{C}_\theta = \left[ \begin{array}{cc}
\cos \theta &  -i\sin \theta \\
-i\sin \theta & \cos \theta
\end{array}\right]\otimes \hat{I}_{p},
\label{cn}
\end{align}
where $\theta$ is a coin bias parameter that can be varied at each time-step, each position or both to implement different kinds of quantum walks, and $\hat{I}_p$ is the identity operator in the position space\,\cite{andraca2012}. The conditional position-shift operator, $\hat{S}$, translates the particle to the left and right conditioned by the state of the coin qubit,
\begin{align}\label{sh}
\hat{S}  = \sum_{x\in\mathbb{Z}}  \Big [ \vert 0 \rangle \langle 0 \vert \otimes  \vert x-1 \rangle \langle x \vert + \vert 1  \rangle \langle 1 \vert \otimes  \vert x+1 \rangle \langle x \vert \Big ].
\end{align}
The state of the particle after $t$ steps of the walk, is obtained by the repeated action of the operator $\hat{W}=\hat{S}\hat{C}_{\theta}$ on the initial state of the particle $\vert \psi \rangle_{c} = \alpha \vert 0 \rangle + \beta \vert 1 \rangle$ (for $\alpha, \beta \in \mathbb{C}$) at position $x=0$,
\begin{align}\label{ev}
\ket{\Psi(x,t)} = \hat{W}^t \bigg[ \ket{\psi}_c \otimes \ket{x=0} \bigg ] = \sum_x \left[\begin{array}{c}
\psi^{l}_{x, t} \\ \psi^{r}_{x, t}
\end{array}\right],
\end{align}
where $\psi_{x,t}^{l(r)}$ denotes the left (right) propagating component of the particle at time-step $t$. The probability of finding the particle at position $x$ and time $t$ will be $P(x, t) = \vert \psi^{l}_{x, t}\vert^2 + \vert\psi^{r}_{x, t}\vert^2$.

\smallskip
{\noindent \bf Split-step quantum walks:} For the ease of understanding of realization scheme, we briefly outline the SSQW as well. Each step of a SSQW is a composition of two half step evolutions with different coin bases and position shift operators,
\begin{align} \label{wss}
\hat{W}_{ss} = \hat{S}_{+}\hat{C}_{\theta_2} \hat{S}_{-} \hat{C}_{\theta_1},
\end{align}
where the coin operation $\hat{C}_{\theta_k}$, with $k=1,2$, is given in Eq.\,(\ref{cn}). We note that in the general case, these coin operators can be chosen as the elements of the U(2) group. However, because the overall phase does not matter, we can, without loss of generality, consider both coin operators to be elements of the group SU(2). The split-step position shift operators are,
\begin{subequations}
	\begin{align} 
	\hat{S}_{-} &=  |0\rangle\langle 0| \otimes \sum_{x\in\mathbb{Z}}  |x-1\rangle\langle x|+|1\rangle\langle 1| \otimes \sum_{x\in\mathbb{Z}}|x\rangle\langle x|, \label{sm} \\
	\hat{S}_{+} &=  |0\rangle\langle 0| \otimes \sum_{x\in\mathbb{Z}}  |x\rangle\langle x|+|1\rangle\langle 1| \otimes \sum_{x\in\mathbb{Z}}|x+1\rangle\langle x|. \label{sp}
	\end{align}
\end{subequations}

The repeated applications of $\hat{W}_{ss}$ on an initial state $\ket{\psi}_c$ brings the particle to a state at time $t$ and position $x$ described by the differential equation~\cite{alderete2020}, 
\begin{align}
\label{te}
\frac{\partial}{\partial t} \left[\begin{array}{c}
\psi^l_{x,t} \\ \psi^r_{x,t} 
\end{array}\right] 
&= \cos \theta_{2} \left[\begin{array}{cc}
\cos \theta_1  & -i \sin \theta_1 \\
i \sin \theta_1  & -\cos \theta_1
\end{array}\right] %
\frac{\partial}{\partial x} \left[\begin{array}{c}
\psi^l_{x,t} \\ \psi^r_{x,t} 
\end{array}\right] \nonumber \\
&\,\,\,+ \left[ \begin{array}{cc}
\cos(\theta_1+\theta_2)-1 & -i \sin(\theta_1+\theta_2) \\
-i \sin(\theta_1+\theta_2) & \cos(\theta_1+\theta_2)-1
\end{array}\right] \nonumber \\
&\,\,\,\times \left[ \begin{array}{c}
\psi^l_{x,t} \\ \psi^r_{x,t} 
\end{array}\right].
\end{align}
By controlling the parameters $\theta_1$ and $\theta_2$, the split-step quantum walk turns into the one-dimensional Dirac equations for massless and massive spin-$1/2$ particles~\cite{mallick2016}.

\smallskip
{\noindent \bf Position-dependent coin operator:} According to \eqref{ev}, the homogeneous quantum walk involves repeated application of a single unitary coin operation $\hat{C}$, thus limiting the available space of unitaries (set of $\hat{W}$) for realizing a generic quantum walk. The constraint is lifted by allowing to choose a different coin operation at each lattice position (or integral orbital angular momentum), thus increasing the set of accessible coin operations as a step towards implementing general walks. A PDC operator $\hat{C}^{(x)} \equiv \hat{C}^{(x)}[\chi(x), \xi(x), \eta(x), \theta(x)]$ hosts rotation angles that depend on the position in the lattice, defined as,
\begin{align} \label{ui3}
\hat{C}^{(x)} &= \sum_x e^{i\chi(x)} e^{i\xi(x) \sigma_2} e^{i\eta(x) \sigma_3} e^{i\theta(x) \sigma_2} \otimes |x \rangle \langle x|. 
\end{align}
The $2 \times 2$ matrix on each position vector $|x \rangle$ is an element of the $SU(2)$ group, whereas $e^{i\chi(x)}$ is an element of the $U(1)$ group. 
Correspondingly, the walk unitary at each step will now be, $\hat{W}^{(x)} = \hat{S}\hat{C}^{(x)}_{\theta}$ and the final state after $t$ steps is $$\ket{\Psi(x,t)} = \hat{W}^{(x)}_t\hat{W}^{(x)}_{t-1}\cdots\hat{W}^{(x)}_{1}\Big[\ket{\psi}_c \otimes \ket{x=0}\Big].$$

\smallskip
{\noindent \bf Generalized split-step quantum walks:} We define the generalized version of split-step walks as the SSQW with PDC operators. Unitary operator corresponding to such a walk is
\begin{align} \label{ui}
\hat{U}^{(x)} = \hat{S}_+ \hat{C}^{(x)}_{2} \hat{S}_- \hat{C}^{(x)}_{1},
\end{align}
where $\hat{S}_\pm$ are the shift operators as given in ~\eqref{sm} and \eqref{sp}, and $\hat{C}^{(x)}_{k}$ are PDC operators for $k=\{1, 2\}$ as given in \eqref{ui3}. This is unlike the case of SSQW since the coin parameters there are fixed for all positions. 

\section{Realization of generalized SSQW} \label{re} 
We remark that while the position basis is realized with the OAM degree of freedom\,(DoF) of the single photon, the coin basis is realized with the polarization DoF of the same. We extensively make use of $J$-plates for simultaneously manipulating both OAM and polarization DoF of a single photon. The Jones matrix corresponding to a $J$-plate is
\begin{align}
\mathbf{J}(\delta_x,\delta_y,\vartheta) &= R_{-\vartheta} 
\begin{bmatrix}
e^{i\delta_x} & 0 \\
0 & e^{i\delta_y}
\end{bmatrix} R_{\vartheta}, \label{j1a}
\end{align}
where $R_\vartheta = e^{-i\vartheta\sigma_2}$, with $\sigma_2=\begin{bmatrix}
0 & -i \\
i & 0
\end{bmatrix}$. Here, $\delta_x$ and $\delta_y$ are, respectively, phase shifts provided by the $J$-plate along $x$- and $y$-directions \cite{devlin2017, mueller2017}, and the diagonal matrix on the RHS of Eq.\,(\ref{j1a}) can be thought of as a `phase shifter' matrix with $\vartheta$ as the rotation angle about its $z$-axis. In other words, each point of the $J$-plate acts like a tiny variable waveplate which can rotate independently about the $z$-axis. For our photonic implementation of quantum walks, we employ the polarization DoF spanned by $\{\ket{H}, \ket{V}\}$ to replace the role of coin basis, while OAM modes denoted by \{$\ket{\ell}$\} will replace the position basis. The OAM mode $|\ell \rangle$ is known to contain a transverse phase profile $e^{i\ell \phi}$, where $\phi=\tan^{-1} (y/x)$ with $(x,y)$ denoting the transverse plane coordinates of the beam\,\cite{allen92}. Evidently, the OAM mode $|\ell \rangle$ carries a longitudinal OAM of $\ell \hbar$ per photon\,\cite{allen92}. In our paper, we denote any beam containing the phase profile $e^{i\ell \phi}$ as $|\ell \rangle$, regardless of the amplitude profile of the same.

\subsection{Split-step operators}

The split-step operators $\hat{S}_-$ and $\hat{S}_+$ from Eqs.\,(\ref{sm}) and (\ref{sp}) assume the following form in the polarization and OAM DoF\,:
\begin{align}
\hat{S}_- &= e^{-i\phi} |H \rangle \langle H| + |V \rangle \langle V|, \label{s-} \\
\hat{S}_+ &= |H \rangle \langle H| + e^{i\phi} |V \rangle \langle V|, \label{s+} 
\end{align}
where $\phi=\tan^{-1} (y/x)$, with $(x,y)$ being the coordinates in the transverse plane in which a $J$-plate is kept. We can implement these split-step operators using dynamic wave retarders\,\cite{saleh2007} which are physically achieved through a spatial light modulator. These split-step operators can also be realized using a $J$-plate as follows. The operator $\hat{S}_-$ is realized using the $J$-plate $\mathbf{J}(-\phi,0,0)$, and $\hat{S}_+$ is realized using the $J$-plate $\mathbf{J}(0,\phi,0)$. Thus, when a horizontal\,(vertical) polarized single photon carrying an OAM of $\ell \hbar$ per photon, namely, the mode $|\ell \rangle$, passes through a $J$-plate implementing the operator $\hat{S}_-$\,($\hat{S}_+$), its OAM gets converted to $|\ell-1 \rangle$\,($|\ell+1 \rangle$); however, its polarization remains unchanged. As an aside, it can be noted that the shift operator $\hat{S}$ described in Eq.\,(\ref{sh}) appearing in the DTQW can be realized with a $J$-plate $\mathbf{J}(-\phi,\phi,0)$.

\begin{figure}[htbp]
\centering
\includegraphics[scale=0.4]{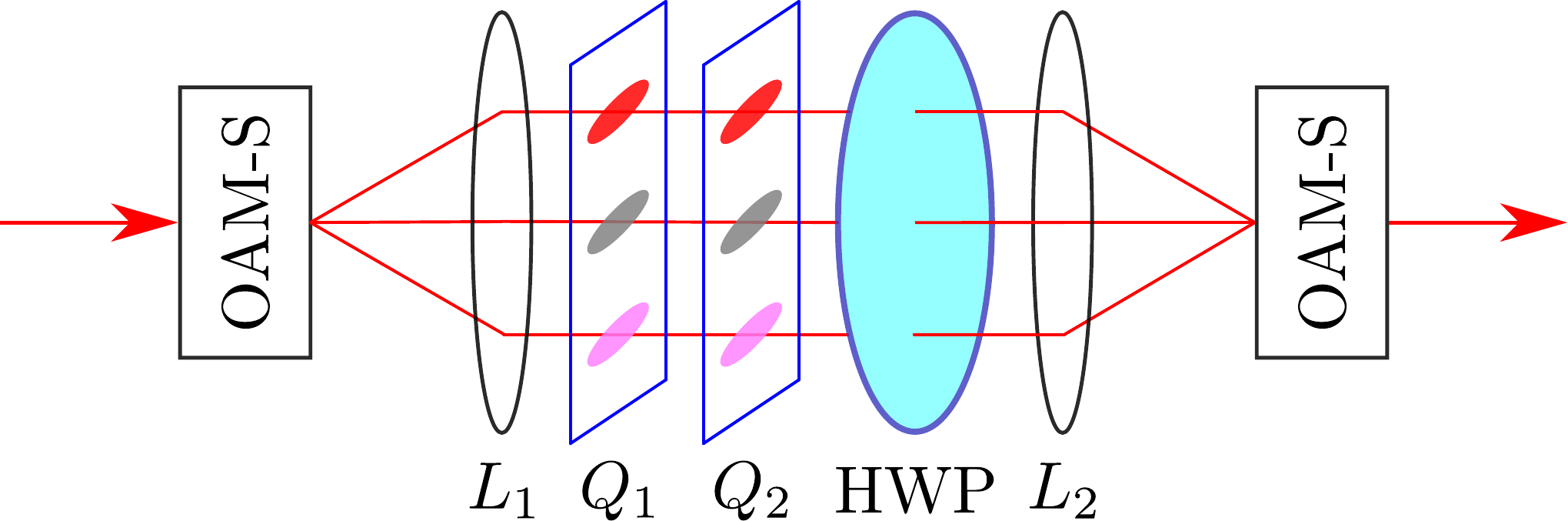}
\caption{A realization scheme for PDC operation, represented by the operator $\hat{C}^{(x)}$\,[see Eq.\,(\ref{cx1})]. The plates $Q_1$, $Q_2$, and half-waveplate\,(HWP) implement the U(2) operation on each OAM mode, where $Q_1$ and $Q_2$ are realized using $J$-plates\,[see Eq.\,(\ref{cx1})]. While the first OAM-sorter\,(OAM-S) spatially separates the constituent OAM modes present in the incoming beam, the second one acts in reverse to recombine them.}
\label{pdc}
\end{figure}

\subsection{Position-dependent coin operator}

\begin{figure}[htbp]
\centering
\includegraphics[scale=0.35]{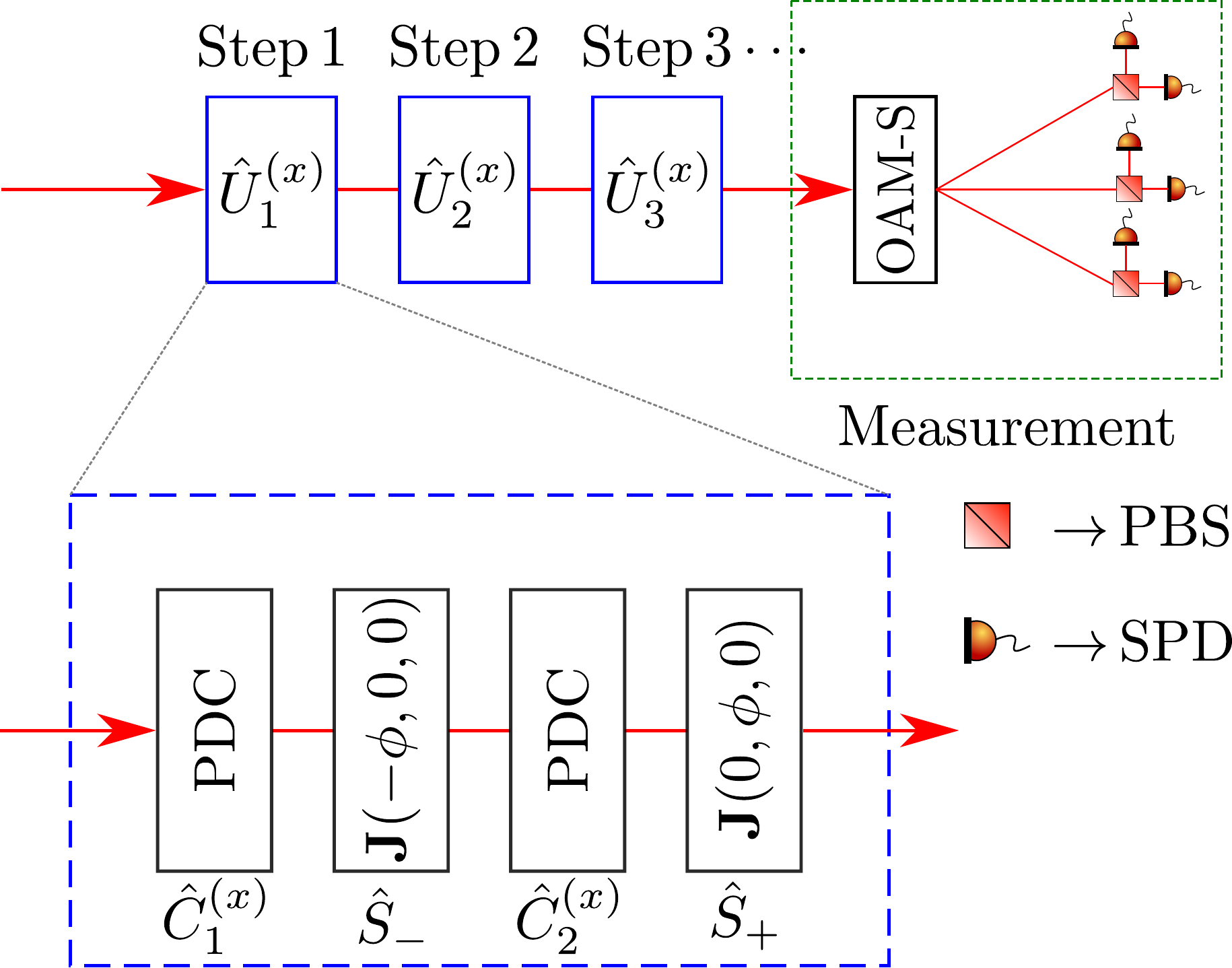}
\caption{A realization scheme for generalized SSQW operation. In each step of the walk, the unitary operations given in Eq.\,(\ref{ui}) are sequentially applied to the incoming single photon. Here, the shift operators, $\hat{S}_-$ and $\hat{S}_+$, are realized using $J$-plates\,[see Eq.\,(\ref{j1a})]. On the other hand, the coin operators $\hat{C}_1^{(x)}$ and $\hat{C}_2^{(x)}$ are realized by position-dependent coin\,(PDC) operations as shown in Fig.\,\ref{pdc}. Finally, measurement of probability amplitudes can be performed using an assembly of OAM sorter~(OAM-S), polarization beamsplitters~(PBS), and single photon detectors~(SPD).}
\label{gssqw}
\end{figure}

A realization scheme for the PDC operator $\hat{C}^{(x)}$ from~\eqref{ui3} is shown in Fig.\,\ref{pdc}. First, we consider a single photon in the state $|\psi \rangle = \sum_{\ell} \big(c_\ell^{(H)} |H \rangle + c_\ell^{(V)} |V \rangle\big) \otimes |\ell \rangle$, where $c_\ell^{(H)}$'s and $c_\ell^{(V)}$'s are normalized coefficients pertaining to a given OAM mode $\ket{\ell}$. In principle, the state $|\psi \rangle$ can be generated using either a dynamic wave retarder\,\cite{saleh2007} or in combination of polarization beamsplitters and holograms\,\cite{heckenberg92}. Using the OAM sorter (OAM-S)\,\cite{berkhout2010, mirhosseini2013}, photons with different OAM values are spatially sorted so that modes with OAM values fall into the lens $L_1$ as $\{\cdots,-1,0,1,\cdots\}$. The modes reaching lens $L_1$ are assumed to be sufficiently separated from each other and do not overlap. By passing through $L_1$, the modes travel parallel to the $z$-axis. Now each of these OAM modes is imparted with a U(2) group operation in their respective polarization state as follows. It can be easily verified that the PDC operator, $\hat{C}^{(x)}$\,[Eq.\,(\ref{ui3})], can be realized using $J$-plates\,[Eq.\,(\ref{j1a})] as
\begin{align} \label{cx1}
\hat{C}^{(x)} &= \sum_x \left( e^{i\xi(x) \sigma_2} 
\begin{bmatrix}
e^{i[\chi(x)+\eta(x)]} & 0 \\
0 & e^{i[\chi(x)-\eta(x)]}
\end{bmatrix} 
e^{-i\xi(x) \sigma_2} \right) \nonumber \\
&\,\,\,\times e^{i[\theta(x)+\xi(x)] \sigma_2} \otimes |x \rangle \langle x| \nonumber \\
&= \sum_x \mathbf{J}(\chi(x)+\eta(x),\chi(x)-\eta(x), \xi(x)) \nonumber \\
&\,\,\,\times \mathbf{J} \left( 0,\pi, \frac{\theta(x)+\xi(x)}{2} \right) \sigma_3 \otimes |x \rangle \langle x|,
\end{align}
where $\sigma_3={\rm diag}\,(1,-1)$. Because $\sigma_3$ can be realized using a HWP, we find that $\hat{C}^{(x)}$ can be realized using 2 $J$-plates and a HWP. In Fig.\,\ref{pdc} we have labeled these two $J$-plates respectively with $Q_1$ and $Q_2$. Therefore, this arrangement effectively implements a $U(2)$ operation on each OAM mode. Because lens $L_2$ ignores the polarization of the incoming beam, the polarization state of the beam is not affected. We remark that since the polarization state of each OAM mode is different, these OAM modes will not interfere by Fresnel-Arago's law\,\cite{barakat93}. Hence, the output of the sorter and the collimating lens would be an array of collimated non-overlapping beams. Thus these OAM modes with different polarization states are recombined by the lens $L_2$ and sent to the second OAM-S which is operated in reverse to obtain a single beam as desired\,\cite{fickler2014}.

\begin{figure}[htbp]
\centering
\includegraphics[scale=0.35]{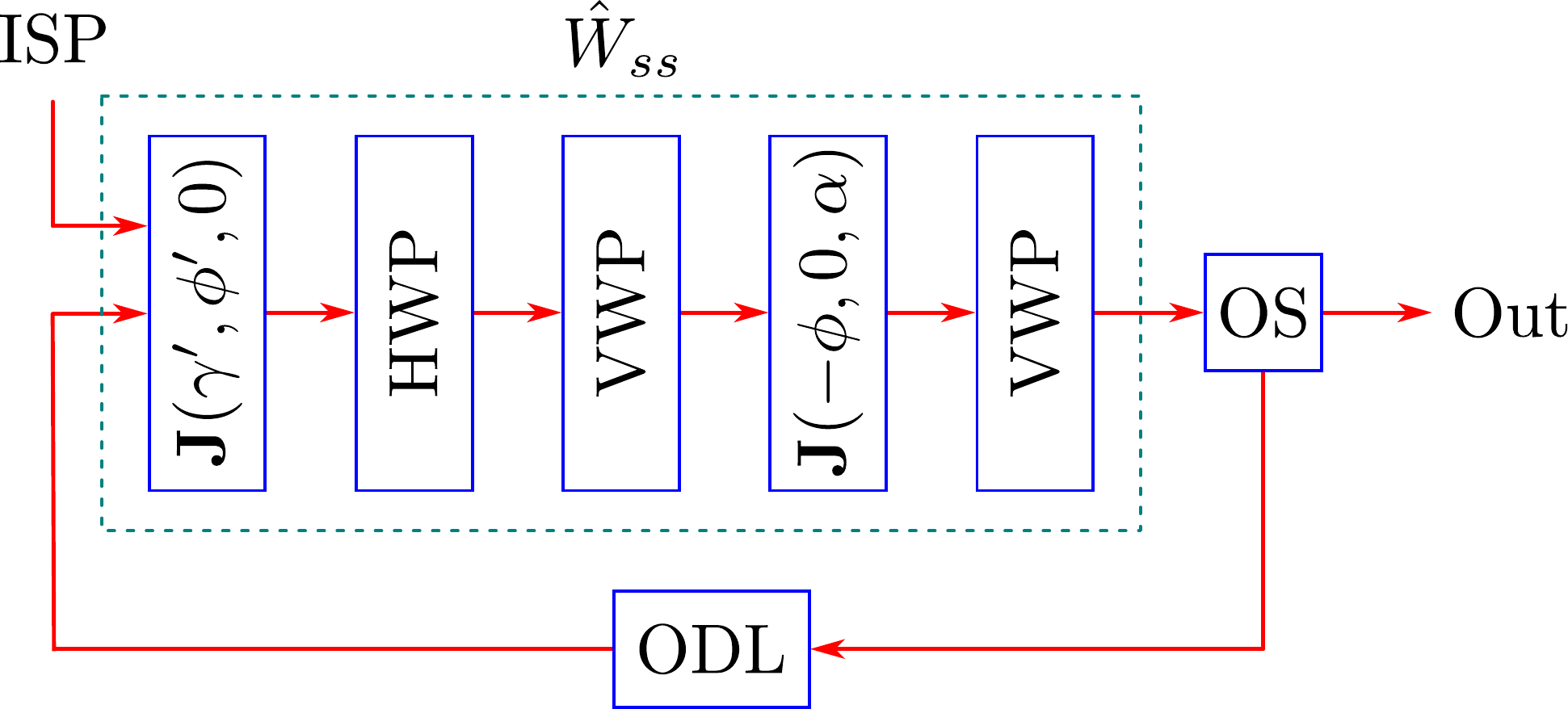}
\caption{Shown here is the optical realization of SSQW. Here, each step operation is realized using 2 $J$-plates, 1 HWP, and 2 variable waveplates\,(VWP) as in Eq.\,(\ref{sr6}), with $\gamma'=(\gamma_2+\pi)/2$ and $\phi'=\phi -(\gamma_1+\pi)/2$. An incoming single photon\,(ISP) passing through these optical elements has performed one step of the SSQW. If the desired number of steps in the SSQW are not performed, the optical switch\,(OS) will make the single photon pass through an optical delay line\,(ODL), and ODL provides the necessary time delay before feeding the photon again to the next step of the SSQW operation. Upon completing the desired number of steps, the OS now lets the single photon through the path labeled `Out'.}
\label{ssqw}
\end{figure}

\subsection{SSQW Realization}

We now show how the SSQW operation can be realized using $J$-plates, variable waveplates, and HWPs. We observe that $\hat{W}_{ss}$ in Eq.\,(\ref{wss}) can be rewritten as
\begin{align} \label{sr1}
\hat{W}_{ss} = \hat{S}_+ \hat{C}_3 (\hat{C}_1^\dagger \hat{S}_- \hat{C}_1),
\end{align}
where $\hat{C}_3=\hat{C}_2 \hat{C}_1$. Suppose $|u_1 \rangle$ and $|u_2 \rangle$ are the column vectors of $\hat{C}_1^\dagger$. Making use of Eq.\,(\ref{s-}) we find that,
\begin{align} \label{sr2}
\hat{C}_1^\dagger \hat{S}_- \hat{C}_1 &= e^{-i\phi} |u_1 \rangle \langle u_1| + |u_2 \rangle \langle u_2|.
\end{align}
Because the column vectors $|u_1 \rangle$ and $|u_2 \rangle$ are orthogonal, we can conveniently parameterize them as,
$\ket{u_1} =  (\cos \alpha,\, e^{i\beta} \sin\alpha)^T$, and $\ket{u_2} = (-\sin\alpha,\, e^{i\beta} \cos\alpha)^T$,
where $\alpha$ and $\beta$ respect $0 \leq \alpha \leq 2\pi$ and $0 \leq \beta <2\pi$. It is now straightforward to verify that
\begin{align} \label{sr4}
\hat{C}_1^\dagger \hat{S}_- \hat{C}_1 = e^{i(\pi-\beta) \sigma_3/2} \mathbf{J}(-\phi,0,\alpha) e^{-i(\pi-\beta) \sigma_3/2},
\end{align}
Availing the Euler decomposition allows the following form for $\hat{C}_3$:
\begin{align} \label{sr5}
\hat{C}_3 &= e^{i\gamma_1 \sigma_3/2} e^{i\gamma_2 \sigma_2/2} e^{i\gamma_3 \sigma_3/2} \nonumber \\
&= (-i) e^{i(\gamma_1+\pi) \sigma_3/2}
\begin{bmatrix}
\cos \gamma_2/2 & \sin \gamma_2/2 \\
\sin \gamma_2/2 & -\cos \gamma_2/2
\end{bmatrix} e^{i\gamma_3 \sigma_3/2}.
\end{align}
Substituting Eqs.\,(\ref{s+}), (\ref{sr4}), and (\ref{sr5}) in Eq.\,(\ref{sr1}) simplifies the walk unitary to an optically amenable form,
\begin{align} \label{sr6}
\hat{W}_{ss} &= (-i) \mathbf{J}\big((\gamma_2+\pi)/2, \phi -(\gamma_1+\pi)/2,0\big) \nonumber \\
&\,\,\,\times 
\begin{bmatrix}
\cos \gamma_2/2 & \sin \gamma_2/2 \\
\sin \gamma_2/2 & -\cos \gamma_2/2
\end{bmatrix} e^{i(\pi-\beta+\gamma_3) \sigma_3/2} \nonumber \\
&\,\,\,\times \mathbf{J}(-\phi,0,\alpha) \, e^{-i(\pi-\beta) \sigma_3/2}.
\end{align}
Ignoring the overall phase factor $(-i)$, we realize each of $\mathbf{J}\big((\gamma_1+\pi)/2, \phi -(\gamma_1+\pi)/2,0\big)$ and $\mathbf{J}(-\phi,0,\alpha)$ using a $J$-plate. Likewise, each of $e^{i(\pi-\beta+\gamma_3) \sigma_3/2}$ and $e^{-i(\pi-\beta) \sigma_3/2}$ can be realized using a variable waveplate. The remaining $2 \times 2$ rotation matrix on the RHS can be realized using a rotated HWP. This just constitutes one step of the SSQW. Making use of an optical switch\,(OS)\,\cite{ono2020,guo2022} and an optical delay line\,(ODL), we can let the single photon to pass through the same set of optical components $t$ times and realize a $t$-step SSQW. This has been illustrated in Fig.\,\ref{ssqw}. We remark that DTQW operation with SU(2) coin can be realized by replacing $\hat{W}_{ss}$ operation in the Fig.\,\ref{ssqw} with $\hat{W}$\,[see Eq.\,(\ref{ev})] which requires a $J$-plate, 2 quarter waveplates, and a HWP. Further, directed quantum walk can also be realized using DTQW setup by appropriately modifying the shift operation\,(or equivalently, the $J$-plate parameters). These three types of quantum walks are equivalent under certain conditions, and the relation between them can be found in Ref.\,\cite{shivani2021}.

\subsection{Generalized SSQW operation} 
In order to achieve the generalized SSQW, the operators $\hat{C}_1^{(x)}$, $\hat{S}_-$, $\hat{C}_2^{(x)}$, and $\hat{S}_+$, as in Eq.\,(\ref{ui}), need to be applied sequentially to the incoming photons. A sequential operation must be repeated at every step of the quantum walk, as illustrated in Fig.\,\ref{gssqw}. Finally, a measurement operation can be performed as follows. By using an OAM sorter, all the constituent OAM modes can be spatially separated. Each OAM mode requires the use of a polarization beamsplitter (PBS) that separates both horizontal and vertical polarization states. In order to sort $N$ OAM modes before measuring, we will need one OAM sorter and $2N$ PBS. Using single photon detectors (SPD), the probability coefficients can be measured.

\section{Discussions and Concluding remarks} \label{di}

Motivated by the expandability and integrability offered by photonic quantum walks, we make use of OAM and polarization DoF in this proposal for implementation of step and coin operators and thus paving way to realization of various walks. We bring forth an assembly of OAM-sorters to realize PDC operations which inserts a $U(1)$ phase to the walk unitary. Remarking that SSQW procures a natural means to engineer Dirac Hamiltonian, we have emphasized the importance for their optical implementation. Using a combination of $J$-plates, variable waveplates and HWPs, we were able to provide a simpler setup for step and position-independent coin operators. In conjunction with OAM-sorters, we added PDC operations to the realization scheme as a way to emulate generalized SSQW. 

One of the possible applications of PDC operator is to implement electric quantum walks. A DTQW can be turned into an electric walk upon acting by the operator $\hat{F}_E := \exp(i\varphi \hat{x})$ at each step of the walk, where $\varphi$ represents the phase imparted by the effective electric field present between two adjacent sites. In effect, the electric walk unitary can be formulated as, $\hat{W}_E = \hat{F}_E\hat{W}$. It has been observed that the position-dependent imparting of phase gives rise to the dynamics which has a striking resemblance to Bloch oscillations of a charged particle in electric field~\cite{MWA13}. A quantum walker can also be subdiffused by means of random rotation angles at each step in PDC operators resulting in Anderson-like localization~\cite{AC22}.

As noted in the Introduction, PDC operation has been realized in both polarization-path DoF\,\cite{kitagawa2012,bian2015, xiao2020} as well as polarization-time bin DoF\,\cite{schreiber2011, barkhofen2017}. In our scheme, the walk evolution is implemented in polarization-OAM DoF and can be thought of an one-shot scheme because PDC operation is realized at each step for all OAM modes at the same time\,(see Fig.\,\ref{pdc}). Further, as opposed to the multipath interferometric schemes such as that of Ref.\,\cite{kitagawa2012,bian2015, xiao2020}, the walk evolution happens in a single beam in our proposed scheme. Therefore, we believe that our setup is scalable and stable. Consequently, unlike interferometric methods, our proposed scheme does not suffer from alignment issue. In addition, our proposed OAM-sorter makes use of log-polar coordinate transformation, which is in-principle an invertible transformation. While our setup is based on OAM-sorters proposed in Refs.\,\cite{berkhout2010, mirhosseini2013}, we believe that the degradation of radial degrees of freedom can be weakened with the possible advancements in OAM sorting such as \,\cite{wei2020,ruffato2018, li2017}. Because of these recent experimental advancements, we believe that our setup can realize position-dependent coin operations in a controllable environment.

The optical setups have the added benefit of not needing ultra-low temperatures for performing quantum walks in contrast to superconducting, cold atoms, or ion-trap systems. While the optical components required for a larger number of steps of the walk might prove to be bulky, our SSQW design uses a more compact setup than earlier suggested approaches. Because both coin and shift operations are the same for each step of the SSQW, we do not need to change the parameters of $J$-plates -- which are just (static) metasurfaces -- for each step of the walk\,(see Fig.\,\ref{ssqw}). Consequently, any $t$-step SSQW is readily realized by passing a single photon repeatedly $t$-times through the optical components realizing $\hat{W}_{ss}$. On the other hand, we need to change the $J$-plate parameters for every step of the generalized SSQW operation, as it involves two PDC operations\,(see Figs.\,\ref{pdc} and \ref{gssqw}). Provided we can make use of dynamic metasurfaces\,\cite{li2019, roy2018} in place of static $J$-plates to realize PDC operations, every step of the generalized SSQW operation can be realized using a `repetitive optical setup' as in Fig.\,\ref{ssqw}. However, such repetitive setups demand nanosecond response times, whereas dynamic metasurfaces based on liquid crystal technology can respond in the order of milliseconds only. And faster switching times are possible by using switching mechanisms such as charge injection into semiconductors, albeit having lower efficiency\,\cite{shirmanesh2020}. Therefore, by engineering such dynamic metasurfaces with nanosecond response times, it is possible to have a repetitive optical setup to realize generalized SSQW operation.

\begin{acknowledgements}
We acknowledge the support from the Office of Principal Scientific Advisor to Government of India, project no. Prn.SA/QSim/2020. One of us (P. A. A. Y.) would like to thank Prof. Andrew Forbes and Prof. Arseniy Kuznetsov on discussions regarding switching time of metasurfaces.
\end{acknowledgements}


\begin{thebibliography}{66}%
\makeatletter
\providecommand \@ifxundefined [1]{%
 \@ifx{#1\undefined}
}%
\providecommand \@ifnum [1]{%
 \ifnum #1\expandafter \@firstoftwo
 \else \expandafter \@secondoftwo
 \fi
}%
\providecommand \@ifx [1]{%
 \ifx #1\expandafter \@firstoftwo
 \else \expandafter \@secondoftwo
 \fi
}%
\providecommand \natexlab [1]{#1}%
\providecommand \enquote  [1]{``#1''}%
\providecommand \bibnamefont  [1]{#1}%
\providecommand \bibfnamefont [1]{#1}%
\providecommand \citenamefont [1]{#1}%
\providecommand \href@noop [0]{\@secondoftwo}%
\providecommand \href [0]{\begingroup \@sanitize@url \@href}%
\providecommand \@href[1]{\@@startlink{#1}\@@href}%
\providecommand \@@href[1]{\endgroup#1\@@endlink}%
\providecommand \@sanitize@url [0]{\catcode `\\12\catcode `\$12\catcode
  `\&12\catcode `\#12\catcode `\^12\catcode `\_12\catcode `\%12\relax}%
\providecommand \@@startlink[1]{}%
\providecommand \@@endlink[0]{}%
\providecommand \url  [0]{\begingroup\@sanitize@url \@url }%
\providecommand \@url [1]{\endgroup\@href {#1}{\urlprefix }}%
\providecommand \urlprefix  [0]{URL }%
\providecommand \Eprint [0]{\href }%
\providecommand \doibase [0]{https://doi.org/}%
\providecommand \selectlanguage [0]{\@gobble}%
\providecommand \bibinfo  [0]{\@secondoftwo}%
\providecommand \bibfield  [0]{\@secondoftwo}%
\providecommand \translation [1]{[#1]}%
\providecommand \BibitemOpen [0]{}%
\providecommand \bibitemStop [0]{}%
\providecommand \bibitemNoStop [0]{.\EOS\space}%
\providecommand \EOS [0]{\spacefactor3000\relax}%
\providecommand \BibitemShut  [1]{\csname bibitem#1\endcsname}%
\let\auto@bib@innerbib\@empty
\bibitem [{\citenamefont {Venegas-Andraca}(2012)}]{andraca2012}%
  \BibitemOpen
  \bibfield  {author} {\bibinfo {author} {\bibfnamefont {S.~E.}\ \bibnamefont
  {Venegas-Andraca}},\ }\bibfield  {title} {\bibinfo {title} {Quantum walks: a
  comprehensive review},\ }\href
  {https://doi.org/https://doi.org/10.1007/s11128-012-0432-5} {\bibfield
  {journal} {\bibinfo  {journal} {Quantum Information Processing}\ }\textbf
  {\bibinfo {volume} {11}},\ \bibinfo {pages} {1015} (\bibinfo {year}
  {2012})}\BibitemShut {NoStop}%
\bibitem [{\citenamefont {Childs}(2009)}]{Chi09}%
  \BibitemOpen
  \bibfield  {author} {\bibinfo {author} {\bibfnamefont {A.~M.}\ \bibnamefont
  {Childs}},\ }\bibfield  {title} {\bibinfo {title} {Universal computation by
  quantum walk},\ }\href {https://doi.org/10.1103/PhysRevLett.102.180501}
  {\bibfield  {journal} {\bibinfo  {journal} {Phys. Rev. Lett.}\ }\textbf
  {\bibinfo {volume} {102}},\ \bibinfo {pages} {180501} (\bibinfo {year}
  {2009})}\BibitemShut {NoStop}%
\bibitem [{\citenamefont {Singh}\ \emph
  {et~al.}(2021{\natexlab{a}})\citenamefont {Singh}, \citenamefont {Chawla},
  \citenamefont {Sarkar},\ and\ \citenamefont {Chandrashekar}}]{SCS21}%
  \BibitemOpen
  \bibfield  {author} {\bibinfo {author} {\bibfnamefont {S.}~\bibnamefont
  {Singh}}, \bibinfo {author} {\bibfnamefont {P.}~\bibnamefont {Chawla}},
  \bibinfo {author} {\bibfnamefont {A.}~\bibnamefont {Sarkar}},\ and\ \bibinfo
  {author} {\bibfnamefont {C.~M.}\ \bibnamefont {Chandrashekar}},\ }\bibfield
  {title} {\bibinfo {title} {Universal quantum computing using single-particle
  discrete-time quantum walk},\ }\href
  {https://doi.org/10.1038/s41598-021-91033-5} {\bibfield  {journal} {\bibinfo
  {journal} {Scientific Reports}\ }\textbf {\bibinfo {volume} {11}},\ \bibinfo
  {pages} {11551} (\bibinfo {year} {2021}{\natexlab{a}})}\BibitemShut {NoStop}%
\bibitem [{\citenamefont {Portugal}(2013)}]{Rportugal13}%
  \BibitemOpen
  \bibfield  {author} {\bibinfo {author} {\bibfnamefont {R.}~\bibnamefont
  {Portugal}},\ }\href
  {https://doi.org/https://doi.org/10.1007/978-3-319-97813-0} {\emph {\bibinfo
  {title} {Quantum Walks and Search Algorithms}}}\ (\bibinfo  {publisher}
  {Springer},\ \bibinfo {year} {2013})\BibitemShut {NoStop}%
\bibitem [{\citenamefont {Sarkar}\ and\ \citenamefont
  {Chandrashekar}(2019)}]{SC19}%
  \BibitemOpen
  \bibfield  {author} {\bibinfo {author} {\bibfnamefont {A.}~\bibnamefont
  {Sarkar}}\ and\ \bibinfo {author} {\bibfnamefont {C.~M.}\ \bibnamefont
  {Chandrashekar}},\ }\bibfield  {title} {\bibinfo {title} {Multi-bit quantum
  random number generation from a single qubit quantum walk},\ }\href
  {https://doi.org/10.1038/s41598-019-48844-4} {\bibfield  {journal} {\bibinfo
  {journal} {Scientific Reports}\ }\textbf {\bibinfo {volume} {9}},\ \bibinfo
  {pages} {12323} (\bibinfo {year} {2019})}\BibitemShut {NoStop}%
\bibitem [{\citenamefont {Kitagawa}\ \emph {et~al.}(2010)\citenamefont
  {Kitagawa}, \citenamefont {Rudner}, \citenamefont {Berg},\ and\ \citenamefont
  {Demler}}]{kitagawa2010}%
  \BibitemOpen
  \bibfield  {author} {\bibinfo {author} {\bibfnamefont {T.}~\bibnamefont
  {Kitagawa}}, \bibinfo {author} {\bibfnamefont {M.~S.}\ \bibnamefont
  {Rudner}}, \bibinfo {author} {\bibfnamefont {E.}~\bibnamefont {Berg}},\ and\
  \bibinfo {author} {\bibfnamefont {E.}~\bibnamefont {Demler}},\ }\bibfield
  {title} {\bibinfo {title} {Exploring topological phases with quantum walks},\
  }\href {https://doi.org/http://dx.doi.org/10.1103/PhysRevA.82.033429}
  {\bibfield  {journal} {\bibinfo  {journal} {Phys. Rev. A}\ }\textbf {\bibinfo
  {volume} {82}},\ \bibinfo {pages} {033429} (\bibinfo {year}
  {2010})}\BibitemShut {NoStop}%
\bibitem [{\citenamefont {Chandrashekar}\ \emph {et~al.}(2015)\citenamefont
  {Chandrashekar}, \citenamefont {Obuse},\ and\ \citenamefont {Busch}}]{COB15}%
  \BibitemOpen
  \bibfield  {author} {\bibinfo {author} {\bibfnamefont {C.~M.}\ \bibnamefont
  {Chandrashekar}}, \bibinfo {author} {\bibfnamefont {H.}~\bibnamefont
  {Obuse}},\ and\ \bibinfo {author} {\bibfnamefont {T.}~\bibnamefont {Busch}},\
  }\href {https://doi.org/10.48550/ARXIV.1502.00436} {\bibinfo {title}
  {Entanglement properties of localized states in 1d topological quantum
  walks}} (\bibinfo {year} {2015})\BibitemShut {NoStop}%
\bibitem [{\citenamefont {D'Errico}\ \emph {et~al.}(2020)\citenamefont
  {D'Errico}, \citenamefont {Cardano}, \citenamefont {Maffei}, \citenamefont
  {Dauphin}, \citenamefont {Barboza}, \citenamefont {Esposito}, \citenamefont
  {Piccirillo}, \citenamefont {Lewenstein}, \citenamefont {Massignan},\ and\
  \citenamefont {Marrucci}}]{errico2020}%
  \BibitemOpen
  \bibfield  {author} {\bibinfo {author} {\bibfnamefont {A.}~\bibnamefont
  {D'Errico}}, \bibinfo {author} {\bibfnamefont {F.}~\bibnamefont {Cardano}},
  \bibinfo {author} {\bibfnamefont {M.}~\bibnamefont {Maffei}}, \bibinfo
  {author} {\bibfnamefont {A.}~\bibnamefont {Dauphin}}, \bibinfo {author}
  {\bibfnamefont {R.}~\bibnamefont {Barboza}}, \bibinfo {author} {\bibfnamefont
  {C.}~\bibnamefont {Esposito}}, \bibinfo {author} {\bibfnamefont
  {B.}~\bibnamefont {Piccirillo}}, \bibinfo {author} {\bibfnamefont
  {M.}~\bibnamefont {Lewenstein}}, \bibinfo {author} {\bibfnamefont
  {P.}~\bibnamefont {Massignan}},\ and\ \bibinfo {author} {\bibfnamefont
  {L.}~\bibnamefont {Marrucci}},\ }\bibfield  {title} {\bibinfo {title}
  {Two-dimensional topological quantum walks in the momentum space of
  structured light},\ }\href
  {https://doi.org/https://doi.org/10.1364/OPTICA.365028} {\bibfield  {journal}
  {\bibinfo  {journal} {Optica}\ }\textbf {\bibinfo {volume} {7}},\ \bibinfo
  {pages} {108} (\bibinfo {year} {2020})}\BibitemShut {NoStop}%
\bibitem [{\citenamefont {Zhang}\ \emph {et~al.}(2017)\citenamefont {Zhang},
  \citenamefont {Goyal}, \citenamefont {Simon},\ and\ \citenamefont
  {Sanders}}]{zhang2017}%
  \BibitemOpen
  \bibfield  {author} {\bibinfo {author} {\bibfnamefont {W.-W.}\ \bibnamefont
  {Zhang}}, \bibinfo {author} {\bibfnamefont {S.~K.}\ \bibnamefont {Goyal}},
  \bibinfo {author} {\bibfnamefont {C.}~\bibnamefont {Simon}},\ and\ \bibinfo
  {author} {\bibfnamefont {B.~C.}\ \bibnamefont {Sanders}},\ }\bibfield
  {title} {\bibinfo {title} {Decomposition of split-step quantum walks for
  simulating majorana modes and edge states},\ }\href
  {https://doi.org/https://doi.org/10.1103/PhysRevA.95.052351} {\bibfield
  {journal} {\bibinfo  {journal} {Phys. Rev. A}\ }\textbf {\bibinfo {volume}
  {95}},\ \bibinfo {pages} {052351} (\bibinfo {year} {2017})}\BibitemShut
  {NoStop}%
\bibitem [{\citenamefont {D’Errico}\ \emph {et~al.}(2021)\citenamefont
  {D’Errico}, \citenamefont {Barboza}, \citenamefont {Tudor}, \citenamefont
  {Dauphin}, \citenamefont {Massignan}, \citenamefont {Marrucci},\ and\
  \citenamefont {Cardano}}]{errico2021}%
  \BibitemOpen
  \bibfield  {author} {\bibinfo {author} {\bibfnamefont {A.}~\bibnamefont
  {D’Errico}}, \bibinfo {author} {\bibfnamefont {R.}~\bibnamefont {Barboza}},
  \bibinfo {author} {\bibfnamefont {R.}~\bibnamefont {Tudor}}, \bibinfo
  {author} {\bibfnamefont {A.}~\bibnamefont {Dauphin}}, \bibinfo {author}
  {\bibfnamefont {P.}~\bibnamefont {Massignan}}, \bibinfo {author}
  {\bibfnamefont {L.}~\bibnamefont {Marrucci}},\ and\ \bibinfo {author}
  {\bibfnamefont {F.}~\bibnamefont {Cardano}},\ }\bibfield  {title} {\bibinfo
  {title} {Bloch--landau--zener dynamics induced by a synthetic field in a
  photonic quantum walk},\ }\href
  {https://doi.org/https://doi.org/10.1063/5.0037327} {\bibfield  {journal}
  {\bibinfo  {journal} {APL Photonics}\ }\textbf {\bibinfo {volume} {6}},\
  \bibinfo {pages} {020802} (\bibinfo {year} {2021})}\BibitemShut {NoStop}%
\bibitem [{\citenamefont {Giordani}\ \emph {et~al.}(2021)\citenamefont
  {Giordani}, \citenamefont {Innocenti}, \citenamefont {Suprano}, \citenamefont
  {Polino}, \citenamefont {Paternostro}, \citenamefont {Spagnolo},
  \citenamefont {Sciarrino},\ and\ \citenamefont {Ferraro}}]{giordani2021}%
  \BibitemOpen
  \bibfield  {author} {\bibinfo {author} {\bibfnamefont {T.}~\bibnamefont
  {Giordani}}, \bibinfo {author} {\bibfnamefont {L.}~\bibnamefont {Innocenti}},
  \bibinfo {author} {\bibfnamefont {A.}~\bibnamefont {Suprano}}, \bibinfo
  {author} {\bibfnamefont {E.}~\bibnamefont {Polino}}, \bibinfo {author}
  {\bibfnamefont {M.}~\bibnamefont {Paternostro}}, \bibinfo {author}
  {\bibfnamefont {N.}~\bibnamefont {Spagnolo}}, \bibinfo {author}
  {\bibfnamefont {F.}~\bibnamefont {Sciarrino}},\ and\ \bibinfo {author}
  {\bibfnamefont {A.}~\bibnamefont {Ferraro}},\ }\bibfield  {title} {\bibinfo
  {title} {Entanglement transfer, accumulation and retrieval via
  quantum-walk-based qubit--qudit dynamics},\ }\href
  {https://doi.org/https://doi.org/10.1088/1367-2630/abdbe1} {\bibfield
  {journal} {\bibinfo  {journal} {New J. Phys.}\ }\textbf {\bibinfo {volume}
  {23}},\ \bibinfo {pages} {023012} (\bibinfo {year} {2021})}\BibitemShut
  {NoStop}%
\bibitem [{\citenamefont {Broome}\ \emph {et~al.}(2010)\citenamefont {Broome},
  \citenamefont {Fedrizzi}, \citenamefont {Lanyon}, \citenamefont {Kassal},
  \citenamefont {Aspuru-Guzik},\ and\ \citenamefont {White}}]{broome2010}%
  \BibitemOpen
  \bibfield  {author} {\bibinfo {author} {\bibfnamefont {M.~A.}\ \bibnamefont
  {Broome}}, \bibinfo {author} {\bibfnamefont {A.}~\bibnamefont {Fedrizzi}},
  \bibinfo {author} {\bibfnamefont {B.~P.}\ \bibnamefont {Lanyon}}, \bibinfo
  {author} {\bibfnamefont {I.}~\bibnamefont {Kassal}}, \bibinfo {author}
  {\bibfnamefont {A.}~\bibnamefont {Aspuru-Guzik}},\ and\ \bibinfo {author}
  {\bibfnamefont {A.~G.}\ \bibnamefont {White}},\ }\bibfield  {title} {\bibinfo
  {title} {Discrete single-photon quantum walks with tunable decoherence},\
  }\href {https://doi.org/https://doi.org/10.1103/PhysRevLett.104.153602}
  {\bibfield  {journal} {\bibinfo  {journal} {Phys. Rev. Lett.}\ }\textbf
  {\bibinfo {volume} {104}},\ \bibinfo {pages} {153602} (\bibinfo {year}
  {2010})}\BibitemShut {NoStop}%
\bibitem [{\citenamefont {Z{\"a}hringer}\ \emph {et~al.}(2010)\citenamefont
  {Z{\"a}hringer}, \citenamefont {Kirchmair}, \citenamefont {Gerritsma},
  \citenamefont {Solano}, \citenamefont {Blatt},\ and\ \citenamefont
  {Roos}}]{zahringer2010}%
  \BibitemOpen
  \bibfield  {author} {\bibinfo {author} {\bibfnamefont {F.}~\bibnamefont
  {Z{\"a}hringer}}, \bibinfo {author} {\bibfnamefont {G.}~\bibnamefont
  {Kirchmair}}, \bibinfo {author} {\bibfnamefont {R.}~\bibnamefont
  {Gerritsma}}, \bibinfo {author} {\bibfnamefont {E.}~\bibnamefont {Solano}},
  \bibinfo {author} {\bibfnamefont {R.}~\bibnamefont {Blatt}},\ and\ \bibinfo
  {author} {\bibfnamefont {C.~F.}\ \bibnamefont {Roos}},\ }\bibfield  {title}
  {\bibinfo {title} {Realization of a quantum walk with one and two trapped
  ions},\ }\href
  {https://doi.org/http://dx.doi.org/10.1103/PhysRevLett.104.100503} {\bibfield
   {journal} {\bibinfo  {journal} {Phys. Rev. Lett.}\ }\textbf {\bibinfo
  {volume} {104}},\ \bibinfo {pages} {100503} (\bibinfo {year}
  {2010})}\BibitemShut {NoStop}%
\bibitem [{\citenamefont {Huerta~Alderete}\ \emph {et~al.}(2020)\citenamefont
  {Huerta~Alderete}, \citenamefont {Singh}, \citenamefont {Nguyen},
  \citenamefont {Zhu}, \citenamefont {Balu}, \citenamefont {Monroe},
  \citenamefont {Chandrashekar},\ and\ \citenamefont {Linke}}]{alderete2020}%
  \BibitemOpen
  \bibfield  {author} {\bibinfo {author} {\bibfnamefont {C.}~\bibnamefont
  {Huerta~Alderete}}, \bibinfo {author} {\bibfnamefont {S.}~\bibnamefont
  {Singh}}, \bibinfo {author} {\bibfnamefont {N.~H.}\ \bibnamefont {Nguyen}},
  \bibinfo {author} {\bibfnamefont {D.}~\bibnamefont {Zhu}}, \bibinfo {author}
  {\bibfnamefont {R.}~\bibnamefont {Balu}}, \bibinfo {author} {\bibfnamefont
  {C.}~\bibnamefont {Monroe}}, \bibinfo {author} {\bibfnamefont
  {C.}~\bibnamefont {Chandrashekar}},\ and\ \bibinfo {author} {\bibfnamefont
  {N.~M.}\ \bibnamefont {Linke}},\ }\bibfield  {title} {\bibinfo {title}
  {Quantum walks and dirac cellular automata on a programmable trapped-ion
  quantum computer},\ }\href
  {https://doi.org/https://doi.org/10.1038/s41467-020-17519-4} {\bibfield
  {journal} {\bibinfo  {journal} {Nat. Commun.}\ }\textbf {\bibinfo {volume}
  {11}},\ \bibinfo {pages} {1} (\bibinfo {year} {2020})}\BibitemShut {NoStop}%
\bibitem [{\citenamefont {Karski}\ \emph {et~al.}(2009)\citenamefont {Karski},
  \citenamefont {F{\"o}rster}, \citenamefont {Choi}, \citenamefont {Steffen},
  \citenamefont {Alt}, \citenamefont {Meschede},\ and\ \citenamefont
  {Widera}}]{karski2009}%
  \BibitemOpen
  \bibfield  {author} {\bibinfo {author} {\bibfnamefont {M.}~\bibnamefont
  {Karski}}, \bibinfo {author} {\bibfnamefont {L.}~\bibnamefont {F{\"o}rster}},
  \bibinfo {author} {\bibfnamefont {J.-M.}\ \bibnamefont {Choi}}, \bibinfo
  {author} {\bibfnamefont {A.}~\bibnamefont {Steffen}}, \bibinfo {author}
  {\bibfnamefont {W.}~\bibnamefont {Alt}}, \bibinfo {author} {\bibfnamefont
  {D.}~\bibnamefont {Meschede}},\ and\ \bibinfo {author} {\bibfnamefont
  {A.}~\bibnamefont {Widera}},\ }\bibfield  {title} {\bibinfo {title} {Quantum
  walk in position space with single optically trapped atoms},\ }\href
  {https://doi.org/https://doi.org/10.1126/science.1174436} {\bibfield
  {journal} {\bibinfo  {journal} {Science}\ }\textbf {\bibinfo {volume}
  {325}},\ \bibinfo {pages} {174} (\bibinfo {year} {2009})}\BibitemShut
  {NoStop}%
\bibitem [{\citenamefont {Manouchehri}\ and\ \citenamefont
  {Wango}(2014)}]{KJ14}%
  \BibitemOpen
  \bibfield  {author} {\bibinfo {author} {\bibfnamefont {K.}~\bibnamefont
  {Manouchehri}}\ and\ \bibinfo {author} {\bibfnamefont {J.}~\bibnamefont
  {Wango}},\ }\href {https://doi.org/https://doi.org/10.1007/978-3-642-36014-5}
  {\emph {\bibinfo {title} {Physical Implementation of Quantum Walks}}}\
  (\bibinfo  {publisher} {Springer Berlin},\ \bibinfo {year}
  {2014})\BibitemShut {NoStop}%
\bibitem [{\citenamefont {Boutari}\ \emph {et~al.}(2016)\citenamefont
  {Boutari}, \citenamefont {Feizpour}, \citenamefont {Barz}, \citenamefont
  {Franco}, \citenamefont {Kim}, \citenamefont {Kolthammer},\ and\
  \citenamefont {Walmsley}}]{BFB16}%
  \BibitemOpen
  \bibfield  {author} {\bibinfo {author} {\bibfnamefont {J.}~\bibnamefont
  {Boutari}}, \bibinfo {author} {\bibfnamefont {A.}~\bibnamefont {Feizpour}},
  \bibinfo {author} {\bibfnamefont {S.}~\bibnamefont {Barz}}, \bibinfo {author}
  {\bibfnamefont {C.~D.}\ \bibnamefont {Franco}}, \bibinfo {author}
  {\bibfnamefont {M.~S.}\ \bibnamefont {Kim}}, \bibinfo {author} {\bibfnamefont
  {W.~S.}\ \bibnamefont {Kolthammer}},\ and\ \bibinfo {author} {\bibfnamefont
  {I.~A.}\ \bibnamefont {Walmsley}},\ }\bibfield  {title} {\bibinfo {title}
  {Large scale quantum walks by means of optical fiber cavities},\ }\href
  {https://doi.org/10.1088/2040-8978/18/9/094007} {\bibfield  {journal}
  {\bibinfo  {journal} {J. Opt.}\ }\textbf {\bibinfo {volume} {18}},\ \bibinfo
  {pages} {094007} (\bibinfo {year} {2016})}\BibitemShut {NoStop}%
\bibitem [{\citenamefont {Aspuru-Guzik}\ and\ \citenamefont
  {Walther}(2012)}]{AW12}%
  \BibitemOpen
  \bibfield  {author} {\bibinfo {author} {\bibfnamefont {A.}~\bibnamefont
  {Aspuru-Guzik}}\ and\ \bibinfo {author} {\bibfnamefont {P.}~\bibnamefont
  {Walther}},\ }\bibfield  {title} {\bibinfo {title} {Photonic quantum
  simulators},\ }\href {https://doi.org/10.1038/nphys2253} {\bibfield
  {journal} {\bibinfo  {journal} {Nature Physics}\ }\textbf {\bibinfo {volume}
  {8}},\ \bibinfo {pages} {285} (\bibinfo {year} {2012})}\BibitemShut {NoStop}%
\bibitem [{\citenamefont {Roldán}\ and\ \citenamefont {Soriano}(2005)}]{EJ05}%
  \BibitemOpen
  \bibfield  {author} {\bibinfo {author} {\bibfnamefont {E.}~\bibnamefont
  {Roldán}}\ and\ \bibinfo {author} {\bibfnamefont {J.~C.}\ \bibnamefont
  {Soriano}},\ }\bibfield  {title} {\bibinfo {title} {Optical implementability
  of the two-dimensional quantum walk},\ }\href
  {https://doi.org/10.1080/09500340500309873} {\bibfield  {journal} {\bibinfo
  {journal} {J. Mod. Opt.}\ }\textbf {\bibinfo {volume} {52}},\ \bibinfo
  {pages} {2649} (\bibinfo {year} {2005})}\BibitemShut {NoStop}%
\bibitem [{\citenamefont {Yasir}\ and\ \citenamefont
  {Chandrashekar}(2022)}]{yasir2022}%
  \BibitemOpen
  \bibfield  {author} {\bibinfo {author} {\bibfnamefont {P.~A.~A.}\
  \bibnamefont {Yasir}}\ and\ \bibinfo {author} {\bibfnamefont {C.~M.}\
  \bibnamefont {Chandrashekar}},\ }\bibfield  {title} {\bibinfo {title}
  {Generation of hyperentangled states and two-dimensional quantum walks using
  j or q plates and polarization beam splitters},\ }\href
  {https://doi.org/https://doi.org/10.1103/PhysRevA.105.012417} {\bibfield
  {journal} {\bibinfo  {journal} {Phys. Rev. A}\ }\textbf {\bibinfo {volume}
  {105}},\ \bibinfo {pages} {012417} (\bibinfo {year} {2022})}\BibitemShut
  {NoStop}%
\bibitem [{\citenamefont {Zhang}\ \emph {et~al.}(2010)\citenamefont {Zhang},
  \citenamefont {Liu}, \citenamefont {Liu}, \citenamefont {Li}, \citenamefont
  {Li},\ and\ \citenamefont {Guo}}]{zhang2010}%
  \BibitemOpen
  \bibfield  {author} {\bibinfo {author} {\bibfnamefont {P.}~\bibnamefont
  {Zhang}}, \bibinfo {author} {\bibfnamefont {B.-H.}\ \bibnamefont {Liu}},
  \bibinfo {author} {\bibfnamefont {R.-F.}\ \bibnamefont {Liu}}, \bibinfo
  {author} {\bibfnamefont {H.-R.}\ \bibnamefont {Li}}, \bibinfo {author}
  {\bibfnamefont {F.-L.}\ \bibnamefont {Li}},\ and\ \bibinfo {author}
  {\bibfnamefont {G.-C.}\ \bibnamefont {Guo}},\ }\bibfield  {title} {\bibinfo
  {title} {Implementation of one-dimensional quantum walks on spin-orbital
  angular momentum space of photons},\ }\href
  {https://doi.org/http://dx.doi.org/10.1103/PhysRevA.81.052322} {\bibfield
  {journal} {\bibinfo  {journal} {Phys. Rev. A}\ }\textbf {\bibinfo {volume}
  {81}},\ \bibinfo {pages} {052322} (\bibinfo {year} {2010})}\BibitemShut
  {NoStop}%
\bibitem [{\citenamefont {Cardano}\ \emph {et~al.}(2015)\citenamefont
  {Cardano}, \citenamefont {Massa}, \citenamefont {Qassim}, \citenamefont
  {Karimi}, \citenamefont {Slussarenko}, \citenamefont {Paparo}, \citenamefont
  {de~Lisio}, \citenamefont {Sciarrino}, \citenamefont {Santamato},
  \citenamefont {Boyd},\ and\ \citenamefont {Marrucci}}]{cardano2015}%
  \BibitemOpen
  \bibfield  {author} {\bibinfo {author} {\bibfnamefont {F.}~\bibnamefont
  {Cardano}}, \bibinfo {author} {\bibfnamefont {F.}~\bibnamefont {Massa}},
  \bibinfo {author} {\bibfnamefont {H.}~\bibnamefont {Qassim}}, \bibinfo
  {author} {\bibfnamefont {E.}~\bibnamefont {Karimi}}, \bibinfo {author}
  {\bibfnamefont {S.}~\bibnamefont {Slussarenko}}, \bibinfo {author}
  {\bibfnamefont {D.}~\bibnamefont {Paparo}}, \bibinfo {author} {\bibfnamefont
  {C.}~\bibnamefont {de~Lisio}}, \bibinfo {author} {\bibfnamefont
  {F.}~\bibnamefont {Sciarrino}}, \bibinfo {author} {\bibfnamefont
  {E.}~\bibnamefont {Santamato}}, \bibinfo {author} {\bibfnamefont {R.~W.}\
  \bibnamefont {Boyd}},\ and\ \bibinfo {author} {\bibfnamefont
  {L.}~\bibnamefont {Marrucci}},\ }\bibfield  {title} {\bibinfo {title}
  {Quantum walks and wavepacket dynamics on a lattice with twisted photons},\
  }\href
  {https://doi.org/https://www.science.org/doi/abs/10.1126/sciadv.1500087}
  {\bibfield  {journal} {\bibinfo  {journal} {Sci. Adv.}\ }\textbf {\bibinfo
  {volume} {1}},\ \bibinfo {pages} {e1500087} (\bibinfo {year} {2015})},\
  \Eprint
  {https://arxiv.org/abs/https://www.science.org/doi/pdf/10.1126/sciadv.1500087}
  {https://www.science.org/doi/pdf/10.1126/sciadv.1500087} \BibitemShut
  {NoStop}%
\bibitem [{\citenamefont {Cardano}\ \emph {et~al.}(2016)\citenamefont
  {Cardano}, \citenamefont {Maffei}, \citenamefont {Massa}, \citenamefont
  {Piccirillo}, \citenamefont {De~Lisio}, \citenamefont {De~Filippis},
  \citenamefont {Cataudella}, \citenamefont {Santamato},\ and\ \citenamefont
  {Marrucci}}]{cardano2016}%
  \BibitemOpen
  \bibfield  {author} {\bibinfo {author} {\bibfnamefont {F.}~\bibnamefont
  {Cardano}}, \bibinfo {author} {\bibfnamefont {M.}~\bibnamefont {Maffei}},
  \bibinfo {author} {\bibfnamefont {F.}~\bibnamefont {Massa}}, \bibinfo
  {author} {\bibfnamefont {B.}~\bibnamefont {Piccirillo}}, \bibinfo {author}
  {\bibfnamefont {C.}~\bibnamefont {De~Lisio}}, \bibinfo {author}
  {\bibfnamefont {G.}~\bibnamefont {De~Filippis}}, \bibinfo {author}
  {\bibfnamefont {V.}~\bibnamefont {Cataudella}}, \bibinfo {author}
  {\bibfnamefont {E.}~\bibnamefont {Santamato}},\ and\ \bibinfo {author}
  {\bibfnamefont {L.}~\bibnamefont {Marrucci}},\ }\bibfield  {title} {\bibinfo
  {title} {Statistical moments of quantum-walk dynamics reveal topological
  quantum transitions},\ }\href
  {https://doi.org/https://doi.org/10.1038/ncomms11439} {\bibfield  {journal}
  {\bibinfo  {journal} {Nat. Commun.}\ }\textbf {\bibinfo {volume} {7}},\
  \bibinfo {pages} {1} (\bibinfo {year} {2016})}\BibitemShut {NoStop}%
\bibitem [{\citenamefont {Cardano}\ \emph {et~al.}(2017)\citenamefont
  {Cardano}, \citenamefont {D’Errico}, \citenamefont {Dauphin}, \citenamefont
  {Maffei}, \citenamefont {Piccirillo}, \citenamefont {de~Lisio}, \citenamefont
  {De~Filippis}, \citenamefont {Cataudella}, \citenamefont {Santamato},
  \citenamefont {Marrucci} \emph {et~al.}}]{cardano2017}%
  \BibitemOpen
  \bibfield  {author} {\bibinfo {author} {\bibfnamefont {F.}~\bibnamefont
  {Cardano}}, \bibinfo {author} {\bibfnamefont {A.}~\bibnamefont {D’Errico}},
  \bibinfo {author} {\bibfnamefont {A.}~\bibnamefont {Dauphin}}, \bibinfo
  {author} {\bibfnamefont {M.}~\bibnamefont {Maffei}}, \bibinfo {author}
  {\bibfnamefont {B.}~\bibnamefont {Piccirillo}}, \bibinfo {author}
  {\bibfnamefont {C.}~\bibnamefont {de~Lisio}}, \bibinfo {author}
  {\bibfnamefont {G.}~\bibnamefont {De~Filippis}}, \bibinfo {author}
  {\bibfnamefont {V.}~\bibnamefont {Cataudella}}, \bibinfo {author}
  {\bibfnamefont {E.}~\bibnamefont {Santamato}}, \bibinfo {author}
  {\bibfnamefont {L.}~\bibnamefont {Marrucci}}, \emph {et~al.},\ }\bibfield
  {title} {\bibinfo {title} {Detection of zak phases and topological invariants
  in a chiral quantum walk of twisted photons},\ }\href
  {https://doi.org/https://doi.org/10.1038/ncomms15516} {\bibfield  {journal}
  {\bibinfo  {journal} {Nat. Commun.}\ }\textbf {\bibinfo {volume} {8}},\
  \bibinfo {pages} {1} (\bibinfo {year} {2017})}\BibitemShut {NoStop}%
\bibitem [{\citenamefont {Wang}\ \emph {et~al.}(2018)\citenamefont {Wang},
  \citenamefont {Chen},\ and\ \citenamefont {Zhang}}]{wang2018}%
  \BibitemOpen
  \bibfield  {author} {\bibinfo {author} {\bibfnamefont {B.}~\bibnamefont
  {Wang}}, \bibinfo {author} {\bibfnamefont {T.}~\bibnamefont {Chen}},\ and\
  \bibinfo {author} {\bibfnamefont {X.}~\bibnamefont {Zhang}},\ }\bibfield
  {title} {\bibinfo {title} {Experimental observation of topologically
  protected bound states with vanishing chern numbers in a two-dimensional
  quantum walk},\ }\href
  {https://doi.org/https://link.aps.org/doi/10.1103/PhysRevLett.121.100501}
  {\bibfield  {journal} {\bibinfo  {journal} {Phys. Rev. Lett.}\ }\textbf
  {\bibinfo {volume} {121}},\ \bibinfo {pages} {100501} (\bibinfo {year}
  {2018})}\BibitemShut {NoStop}%
\bibitem [{\citenamefont {Sephton}\ \emph {et~al.}(2019)\citenamefont
  {Sephton}, \citenamefont {Dudley}, \citenamefont {Ruffato}, \citenamefont
  {Romanato}, \citenamefont {Marrucci}, \citenamefont {Padgett}, \citenamefont
  {Goyal}, \citenamefont {Roux}, \citenamefont {Konrad},\ and\ \citenamefont
  {Forbes}}]{sephton2019}%
  \BibitemOpen
  \bibfield  {author} {\bibinfo {author} {\bibfnamefont {B.}~\bibnamefont
  {Sephton}}, \bibinfo {author} {\bibfnamefont {A.}~\bibnamefont {Dudley}},
  \bibinfo {author} {\bibfnamefont {G.}~\bibnamefont {Ruffato}}, \bibinfo
  {author} {\bibfnamefont {F.}~\bibnamefont {Romanato}}, \bibinfo {author}
  {\bibfnamefont {L.}~\bibnamefont {Marrucci}}, \bibinfo {author}
  {\bibfnamefont {M.}~\bibnamefont {Padgett}}, \bibinfo {author} {\bibfnamefont
  {S.}~\bibnamefont {Goyal}}, \bibinfo {author} {\bibfnamefont
  {F.}~\bibnamefont {Roux}}, \bibinfo {author} {\bibfnamefont {T.}~\bibnamefont
  {Konrad}},\ and\ \bibinfo {author} {\bibfnamefont {A.}~\bibnamefont
  {Forbes}},\ }\bibfield  {title} {\bibinfo {title} {A versatile quantum walk
  resonator with bright classical light},\ }\href
  {https://doi.org/https://doi.org/10.1371/journal.pone.0214891} {\bibfield
  {journal} {\bibinfo  {journal} {PloS one}\ }\textbf {\bibinfo {volume}
  {14}},\ \bibinfo {pages} {e0214891} (\bibinfo {year} {2019})}\BibitemShut
  {NoStop}%
\bibitem [{\citenamefont {Giordani}\ \emph {et~al.}(2019)\citenamefont
  {Giordani}, \citenamefont {Polino}, \citenamefont {Emiliani}, \citenamefont
  {Suprano}, \citenamefont {Innocenti}, \citenamefont {Majury}, \citenamefont
  {Marrucci}, \citenamefont {Paternostro}, \citenamefont {Ferraro},
  \citenamefont {Spagnolo},\ and\ \citenamefont {Sciarrino}}]{giordani2019}%
  \BibitemOpen
  \bibfield  {author} {\bibinfo {author} {\bibfnamefont {T.}~\bibnamefont
  {Giordani}}, \bibinfo {author} {\bibfnamefont {E.}~\bibnamefont {Polino}},
  \bibinfo {author} {\bibfnamefont {S.}~\bibnamefont {Emiliani}}, \bibinfo
  {author} {\bibfnamefont {A.}~\bibnamefont {Suprano}}, \bibinfo {author}
  {\bibfnamefont {L.}~\bibnamefont {Innocenti}}, \bibinfo {author}
  {\bibfnamefont {H.}~\bibnamefont {Majury}}, \bibinfo {author} {\bibfnamefont
  {L.}~\bibnamefont {Marrucci}}, \bibinfo {author} {\bibfnamefont
  {M.}~\bibnamefont {Paternostro}}, \bibinfo {author} {\bibfnamefont
  {A.}~\bibnamefont {Ferraro}}, \bibinfo {author} {\bibfnamefont
  {N.}~\bibnamefont {Spagnolo}},\ and\ \bibinfo {author} {\bibfnamefont
  {F.}~\bibnamefont {Sciarrino}},\ }\bibfield  {title} {\bibinfo {title}
  {Experimental engineering of arbitrary qudit states with discrete-time
  quantum walks},\ }\href
  {https://doi.org/https://link.aps.org/doi/10.1103/PhysRevLett.122.020503}
  {\bibfield  {journal} {\bibinfo  {journal} {Phys. Rev. Lett.}\ }\textbf
  {\bibinfo {volume} {122}},\ \bibinfo {pages} {020503} (\bibinfo {year}
  {2019})}\BibitemShut {NoStop}%
\bibitem [{\citenamefont {Ashwin}\ and\ \citenamefont {Ashvin}(2000)}]{NV00}%
  \BibitemOpen
  \bibfield  {author} {\bibinfo {author} {\bibfnamefont {N.}~\bibnamefont
  {Ashwin}}\ and\ \bibinfo {author} {\bibfnamefont {V.}~\bibnamefont
  {Ashvin}},\ }\href {https://dl.acm.org/doi/10.5555/868264} {\emph {\bibinfo
  {title} {Quantum Walk on the Line}}},\ \bibinfo {type} {Tech. Rep.}\
  (\bibinfo {year} {2000})\ \bibinfo {note}
  {arXiv:quant-ph/0010117}\BibitemShut {NoStop}%
\bibitem [{\citenamefont {Chandrashekar}\ \emph {et~al.}(2008)\citenamefont
  {Chandrashekar}, \citenamefont {Srikanth},\ and\ \citenamefont
  {Laflamme}}]{CSL08}%
  \BibitemOpen
  \bibfield  {author} {\bibinfo {author} {\bibfnamefont {C.~M.}\ \bibnamefont
  {Chandrashekar}}, \bibinfo {author} {\bibfnamefont {R.}~\bibnamefont
  {Srikanth}},\ and\ \bibinfo {author} {\bibfnamefont {R.}~\bibnamefont
  {Laflamme}},\ }\bibfield  {title} {\bibinfo {title} {Optimizing the discrete
  time quantum walk using a su(2) coin},\ }\href
  {https://doi.org/10.1103/PhysRevA.77.032326} {\bibfield  {journal} {\bibinfo
  {journal} {Phys. Rev. A}\ }\textbf {\bibinfo {volume} {77}},\ \bibinfo
  {pages} {032326} (\bibinfo {year} {2008})}\BibitemShut {NoStop}%
\bibitem [{\citenamefont {Chandrashekar}(2012)}]{Cha12}%
  \BibitemOpen
  \bibfield  {author} {\bibinfo {author} {\bibfnamefont {C.~M.}\ \bibnamefont
  {Chandrashekar}},\ }\href {https://doi.org/10.48550/ARXIV.1212.5984}
  {\bibinfo {title} {Disorder induced localization and enhancement of
  entanglement in one- and two-dimensional quantum walks}} (\bibinfo {year}
  {2012})\BibitemShut {NoStop}%
\bibitem [{\citenamefont {Panahiyan}\ and\ \citenamefont
  {Fritzsche}(2018)}]{PF18}%
  \BibitemOpen
  \bibfield  {author} {\bibinfo {author} {\bibfnamefont {S.}~\bibnamefont
  {Panahiyan}}\ and\ \bibinfo {author} {\bibfnamefont {S.}~\bibnamefont
  {Fritzsche}},\ }\bibfield  {title} {\bibinfo {title} {Controlling quantum
  random walk with a step-dependent coin},\ }\href
  {https://doi.org/10.1088/1367-2630/aad899} {\bibfield  {journal} {\bibinfo
  {journal} {New J. Phys.}\ }\textbf {\bibinfo {volume} {20}},\ \bibinfo
  {pages} {083028} (\bibinfo {year} {2018})}\BibitemShut {NoStop}%
\bibitem [{\citenamefont {Kitagawa}\ \emph {et~al.}(2012)\citenamefont
  {Kitagawa}, \citenamefont {Broome}, \citenamefont {Fedrizzi}, \citenamefont
  {Rudner}, \citenamefont {Berg}, \citenamefont {Kassal}, \citenamefont
  {Aspuru-Guzik}, \citenamefont {Demler},\ and\ \citenamefont
  {White}}]{kitagawa2012}%
  \BibitemOpen
  \bibfield  {author} {\bibinfo {author} {\bibfnamefont {T.}~\bibnamefont
  {Kitagawa}}, \bibinfo {author} {\bibfnamefont {M.~A.}\ \bibnamefont
  {Broome}}, \bibinfo {author} {\bibfnamefont {A.}~\bibnamefont {Fedrizzi}},
  \bibinfo {author} {\bibfnamefont {M.~S.}\ \bibnamefont {Rudner}}, \bibinfo
  {author} {\bibfnamefont {E.}~\bibnamefont {Berg}}, \bibinfo {author}
  {\bibfnamefont {I.}~\bibnamefont {Kassal}}, \bibinfo {author} {\bibfnamefont
  {A.}~\bibnamefont {Aspuru-Guzik}}, \bibinfo {author} {\bibfnamefont
  {E.}~\bibnamefont {Demler}},\ and\ \bibinfo {author} {\bibfnamefont {A.~G.}\
  \bibnamefont {White}},\ }\bibfield  {title} {\bibinfo {title} {Observation of
  topologically protected bound states in photonic quantum walks},\ }\href
  {https://doi.org/https://doi.org/10.1038/ncomms1872} {\bibfield  {journal}
  {\bibinfo  {journal} {Nat. Commun.}\ }\textbf {\bibinfo {volume} {3}},\
  \bibinfo {pages} {1} (\bibinfo {year} {2012})}\BibitemShut {NoStop}%
\bibitem [{\citenamefont {Barkhofen}\ \emph {et~al.}(2017)\citenamefont
  {Barkhofen}, \citenamefont {Nitsche}, \citenamefont {Elster}, \citenamefont
  {Lorz}, \citenamefont {G\'abris}, \citenamefont {Jex},\ and\ \citenamefont
  {Silberhorn}}]{barkhofen2017}%
  \BibitemOpen
  \bibfield  {author} {\bibinfo {author} {\bibfnamefont {S.}~\bibnamefont
  {Barkhofen}}, \bibinfo {author} {\bibfnamefont {T.}~\bibnamefont {Nitsche}},
  \bibinfo {author} {\bibfnamefont {F.}~\bibnamefont {Elster}}, \bibinfo
  {author} {\bibfnamefont {L.}~\bibnamefont {Lorz}}, \bibinfo {author}
  {\bibfnamefont {A.}~\bibnamefont {G\'abris}}, \bibinfo {author}
  {\bibfnamefont {I.}~\bibnamefont {Jex}},\ and\ \bibinfo {author}
  {\bibfnamefont {C.}~\bibnamefont {Silberhorn}},\ }\bibfield  {title}
  {\bibinfo {title} {Measuring topological invariants in disordered
  discrete-time quantum walks},\ }\href
  {https://doi.org/https://link.aps.org/doi/10.1103/PhysRevA.96.033846}
  {\bibfield  {journal} {\bibinfo  {journal} {Phys. Rev. A}\ }\textbf {\bibinfo
  {volume} {96}},\ \bibinfo {pages} {033846} (\bibinfo {year}
  {2017})}\BibitemShut {NoStop}%
\bibitem [{\citenamefont {Nitsche}\ \emph {et~al.}(2019)\citenamefont
  {Nitsche}, \citenamefont {Geib}, \citenamefont {Stahl}, \citenamefont {Lorz},
  \citenamefont {Cedzich}, \citenamefont {Barkhofen}, \citenamefont {Werner},\
  and\ \citenamefont {Silberhorn}}]{nitsche2019}%
  \BibitemOpen
  \bibfield  {author} {\bibinfo {author} {\bibfnamefont {T.}~\bibnamefont
  {Nitsche}}, \bibinfo {author} {\bibfnamefont {T.}~\bibnamefont {Geib}},
  \bibinfo {author} {\bibfnamefont {C.}~\bibnamefont {Stahl}}, \bibinfo
  {author} {\bibfnamefont {L.}~\bibnamefont {Lorz}}, \bibinfo {author}
  {\bibfnamefont {C.}~\bibnamefont {Cedzich}}, \bibinfo {author} {\bibfnamefont
  {S.}~\bibnamefont {Barkhofen}}, \bibinfo {author} {\bibfnamefont {R.~F.}\
  \bibnamefont {Werner}},\ and\ \bibinfo {author} {\bibfnamefont
  {C.}~\bibnamefont {Silberhorn}},\ }\bibfield  {title} {\bibinfo {title}
  {Eigenvalue measurement of topologically protected edge states in split-step
  quantum walks},\ }\href
  {https://doi.org/https://dx.doi.org/10.1088/1367-2630/ab12fa} {\bibfield
  {journal} {\bibinfo  {journal} {New J. Phys.}\ }\textbf {\bibinfo {volume}
  {21}},\ \bibinfo {pages} {043031} (\bibinfo {year} {2019})}\BibitemShut
  {NoStop}%
\bibitem [{\citenamefont {Ahmad}\ \emph {et~al.}(2020)\citenamefont {Ahmad},
  \citenamefont {Sajjad},\ and\ \citenamefont {Sajid}}]{ASS20}%
  \BibitemOpen
  \bibfield  {author} {\bibinfo {author} {\bibfnamefont {R.}~\bibnamefont
  {Ahmad}}, \bibinfo {author} {\bibfnamefont {U.}~\bibnamefont {Sajjad}},\ and\
  \bibinfo {author} {\bibfnamefont {M.}~\bibnamefont {Sajid}},\ }\bibfield
  {title} {\bibinfo {title} {One-dimensional quantum walks with a
  position-dependent coin},\ }\href {https://doi.org/10.1088/1572-9494/ab7ec5}
  {\bibfield  {journal} {\bibinfo  {journal} {Communications in Theoretical
  Physics}\ }\textbf {\bibinfo {volume} {72}},\ \bibinfo {pages} {065101}
  (\bibinfo {year} {2020})}\BibitemShut {NoStop}%
\bibitem [{\citenamefont {Bian}\ \emph {et~al.}(2015)\citenamefont {Bian},
  \citenamefont {Li}, \citenamefont {Qin}, \citenamefont {Zhan}, \citenamefont
  {Zhang}, \citenamefont {Sanders},\ and\ \citenamefont {Xue}}]{bian2015}%
  \BibitemOpen
  \bibfield  {author} {\bibinfo {author} {\bibfnamefont {Z.}~\bibnamefont
  {Bian}}, \bibinfo {author} {\bibfnamefont {J.}~\bibnamefont {Li}}, \bibinfo
  {author} {\bibfnamefont {H.}~\bibnamefont {Qin}}, \bibinfo {author}
  {\bibfnamefont {X.}~\bibnamefont {Zhan}}, \bibinfo {author} {\bibfnamefont
  {R.}~\bibnamefont {Zhang}}, \bibinfo {author} {\bibfnamefont {B.~C.}\
  \bibnamefont {Sanders}},\ and\ \bibinfo {author} {\bibfnamefont
  {P.}~\bibnamefont {Xue}},\ }\bibfield  {title} {\bibinfo {title} {Realization
  of single-qubit positive-operator-valued measurement via a one-dimensional
  photonic quantum walk},\ }\href
  {https://doi.org/https://link.aps.org/doi/10.1103/PhysRevLett.114.203602}
  {\bibfield  {journal} {\bibinfo  {journal} {Phys. Rev. Lett.}\ }\textbf
  {\bibinfo {volume} {114}},\ \bibinfo {pages} {203602} (\bibinfo {year}
  {2015})}\BibitemShut {NoStop}%
\bibitem [{\citenamefont {Xiao}\ \emph {et~al.}(2020)\citenamefont {Xiao},
  \citenamefont {Deng}, \citenamefont {Wang}, \citenamefont {Zhu},
  \citenamefont {Wang}, \citenamefont {Yi},\ and\ \citenamefont
  {Xue}}]{xiao2020}%
  \BibitemOpen
  \bibfield  {author} {\bibinfo {author} {\bibfnamefont {L.}~\bibnamefont
  {Xiao}}, \bibinfo {author} {\bibfnamefont {T.}~\bibnamefont {Deng}}, \bibinfo
  {author} {\bibfnamefont {K.}~\bibnamefont {Wang}}, \bibinfo {author}
  {\bibfnamefont {G.}~\bibnamefont {Zhu}}, \bibinfo {author} {\bibfnamefont
  {Z.}~\bibnamefont {Wang}}, \bibinfo {author} {\bibfnamefont {W.}~\bibnamefont
  {Yi}},\ and\ \bibinfo {author} {\bibfnamefont {P.}~\bibnamefont {Xue}},\
  }\bibfield  {title} {\bibinfo {title} {Non-hermitian bulk--boundary
  correspondence in quantum dynamics},\ }\href
  {https://doi.org/https://doi.org/10.1038/s41567-020-0836-6} {\bibfield
  {journal} {\bibinfo  {journal} {Nat. Phys.}\ }\textbf {\bibinfo {volume}
  {16}},\ \bibinfo {pages} {761} (\bibinfo {year} {2020})}\BibitemShut
  {NoStop}%
\bibitem [{\citenamefont {Schreiber}\ \emph {et~al.}(2011)\citenamefont
  {Schreiber}, \citenamefont {Cassemiro}, \citenamefont
  {Poto\ifmmode~\check{c}\else \v{c}\fi{}ek}, \citenamefont {G\'abris},
  \citenamefont {Jex},\ and\ \citenamefont {Silberhorn}}]{schreiber2011}%
  \BibitemOpen
  \bibfield  {author} {\bibinfo {author} {\bibfnamefont {A.}~\bibnamefont
  {Schreiber}}, \bibinfo {author} {\bibfnamefont {K.~N.}\ \bibnamefont
  {Cassemiro}}, \bibinfo {author} {\bibfnamefont {V.}~\bibnamefont
  {Poto\ifmmode~\check{c}\else \v{c}\fi{}ek}}, \bibinfo {author} {\bibfnamefont
  {A.}~\bibnamefont {G\'abris}}, \bibinfo {author} {\bibfnamefont
  {I.}~\bibnamefont {Jex}},\ and\ \bibinfo {author} {\bibfnamefont
  {C.}~\bibnamefont {Silberhorn}},\ }\bibfield  {title} {\bibinfo {title}
  {Decoherence and disorder in quantum walks: From ballistic spread to
  localization},\ }\href
  {https://doi.org/https://link.aps.org/doi/10.1103/PhysRevLett.106.180403}
  {\bibfield  {journal} {\bibinfo  {journal} {Phys. Rev. Lett.}\ }\textbf
  {\bibinfo {volume} {106}},\ \bibinfo {pages} {180403} (\bibinfo {year}
  {2011})}\BibitemShut {NoStop}%
\bibitem [{\citenamefont {Panahiyan}\ and\ \citenamefont
  {Fritzsche}(2021)}]{PF21}%
  \BibitemOpen
  \bibfield  {author} {\bibinfo {author} {\bibfnamefont {S.}~\bibnamefont
  {Panahiyan}}\ and\ \bibinfo {author} {\bibfnamefont {S.}~\bibnamefont
  {Fritzsche}},\ }\bibfield  {title} {\bibinfo {title} {Toward simulation of
  topological phenomena with one-, two-, and three-dimensional quantum walks},\
  }\href {https://doi.org/10.1103/PhysRevA.103.012201} {\bibfield  {journal}
  {\bibinfo  {journal} {Phys. Rev. A}\ }\textbf {\bibinfo {volume} {103}},\
  \bibinfo {pages} {012201} (\bibinfo {year} {2021})}\BibitemShut {NoStop}%
\bibitem [{\citenamefont {Chakraborty}\ \emph {et~al.}(2017)\citenamefont
  {Chakraborty}, \citenamefont {Das}, \citenamefont {Mallick},\ and\
  \citenamefont {Chandrashekar}}]{CDM17}%
  \BibitemOpen
  \bibfield  {author} {\bibinfo {author} {\bibfnamefont {S.}~\bibnamefont
  {Chakraborty}}, \bibinfo {author} {\bibfnamefont {A.}~\bibnamefont {Das}},
  \bibinfo {author} {\bibfnamefont {A.}~\bibnamefont {Mallick}},\ and\ \bibinfo
  {author} {\bibfnamefont {C.~M.}\ \bibnamefont {Chandrashekar}},\ }\bibfield
  {title} {\bibinfo {title} {Quantum ratchet in disordered quantum walk},\
  }\href {https://doi.org/https://doi.org/10.1002/andp.201600346} {\bibfield
  {journal} {\bibinfo  {journal} {Annalen der Physik}\ }\textbf {\bibinfo
  {volume} {529}},\ \bibinfo {pages} {1600346} (\bibinfo {year}
  {2017})}\BibitemShut {NoStop}%
\bibitem [{\citenamefont {Kurzy\ifmmode~\acute{n}\else \'{n}\fi{}ski}\ and\
  \citenamefont {W\'ojcik}(2013)}]{kurzynski2013}%
  \BibitemOpen
  \bibfield  {author} {\bibinfo {author} {\bibfnamefont {P.}~\bibnamefont
  {Kurzy\ifmmode~\acute{n}\else \'{n}\fi{}ski}}\ and\ \bibinfo {author}
  {\bibfnamefont {A.}~\bibnamefont {W\'ojcik}},\ }\bibfield  {title} {\bibinfo
  {title} {Quantum walk as a generalized measuring device},\ }\href
  {https://doi.org/10.1103/PhysRevLett.110.200404} {\bibfield  {journal}
  {\bibinfo  {journal} {Phys. Rev. Lett.}\ }\textbf {\bibinfo {volume} {110}},\
  \bibinfo {pages} {200404} (\bibinfo {year} {2013})}\BibitemShut {NoStop}%
\bibitem [{\citenamefont {Lovett}\ \emph {et~al.}(2010)\citenamefont {Lovett},
  \citenamefont {Cooper}, \citenamefont {Everitt}, \citenamefont {Trevers},\
  and\ \citenamefont {Kendon}}]{lovett2010}%
  \BibitemOpen
  \bibfield  {author} {\bibinfo {author} {\bibfnamefont {N.~B.}\ \bibnamefont
  {Lovett}}, \bibinfo {author} {\bibfnamefont {S.}~\bibnamefont {Cooper}},
  \bibinfo {author} {\bibfnamefont {M.}~\bibnamefont {Everitt}}, \bibinfo
  {author} {\bibfnamefont {M.}~\bibnamefont {Trevers}},\ and\ \bibinfo {author}
  {\bibfnamefont {V.}~\bibnamefont {Kendon}},\ }\bibfield  {title} {\bibinfo
  {title} {Universal quantum computation using the discrete-time quantum
  walk},\ }\href {https://doi.org/http://dx.doi.org/10.1103/PhysRevA.81.042330}
  {\bibfield  {journal} {\bibinfo  {journal} {Phys. Rev. A}\ }\textbf {\bibinfo
  {volume} {81}},\ \bibinfo {pages} {042330} (\bibinfo {year}
  {2010})}\BibitemShut {NoStop}%
\bibitem [{\citenamefont {Arrighi}\ \emph {et~al.}(2016)\citenamefont
  {Arrighi}, \citenamefont {Facchini},\ and\ \citenamefont {Forets}}]{AFF16}%
  \BibitemOpen
  \bibfield  {author} {\bibinfo {author} {\bibfnamefont {P.}~\bibnamefont
  {Arrighi}}, \bibinfo {author} {\bibfnamefont {S.}~\bibnamefont {Facchini}},\
  and\ \bibinfo {author} {\bibfnamefont {M.}~\bibnamefont {Forets}},\
  }\bibfield  {title} {\bibinfo {title} {Quantum walking in curved spacetime},\
  }\href {https://doi.org/10.1007/s11128-016-1335-7} {\bibfield  {journal}
  {\bibinfo  {journal} {Quantum Information Processing}\ }\textbf {\bibinfo
  {volume} {15}},\ \bibinfo {pages} {3467} (\bibinfo {year}
  {2016})}\BibitemShut {NoStop}%
\bibitem [{\citenamefont {Mallick}\ \emph {et~al.}(2019)\citenamefont
  {Mallick}, \citenamefont {Mandal}, \citenamefont {Karan},\ and\ \citenamefont
  {Chandrashekar}}]{mallick2019}%
  \BibitemOpen
  \bibfield  {author} {\bibinfo {author} {\bibfnamefont {A.}~\bibnamefont
  {Mallick}}, \bibinfo {author} {\bibfnamefont {S.}~\bibnamefont {Mandal}},
  \bibinfo {author} {\bibfnamefont {A.}~\bibnamefont {Karan}},\ and\ \bibinfo
  {author} {\bibfnamefont {C.~M.}\ \bibnamefont {Chandrashekar}},\ }\bibfield
  {title} {\bibinfo {title} {Simulating dirac hamiltonian in curved space-time
  by split-step quantum walk},\ }\href
  {https://doi.org/https://doi.org/10.1088/2399-6528/aafe2f} {\bibfield
  {journal} {\bibinfo  {journal} {J. Phys. Commun.}\ }\textbf {\bibinfo
  {volume} {3}},\ \bibinfo {pages} {015012} (\bibinfo {year}
  {2019})}\BibitemShut {NoStop}%
\bibitem [{\citenamefont {Cedzich}\ \emph {et~al.}(2013)\citenamefont
  {Cedzich}, \citenamefont {Ryb{\'a}r}, \citenamefont {Werner}, \citenamefont
  {Alberti}, \citenamefont {Genske},\ and\ \citenamefont
  {Werner}}]{cedzich2013}%
  \BibitemOpen
  \bibfield  {author} {\bibinfo {author} {\bibfnamefont {C.}~\bibnamefont
  {Cedzich}}, \bibinfo {author} {\bibfnamefont {T.}~\bibnamefont {Ryb{\'a}r}},
  \bibinfo {author} {\bibfnamefont {A.~H.}\ \bibnamefont {Werner}}, \bibinfo
  {author} {\bibfnamefont {A.}~\bibnamefont {Alberti}}, \bibinfo {author}
  {\bibfnamefont {M.}~\bibnamefont {Genske}},\ and\ \bibinfo {author}
  {\bibfnamefont {R.~F.}\ \bibnamefont {Werner}},\ }\bibfield  {title}
  {\bibinfo {title} {Propagation of quantum walks in electric fields},\ }\href
  {https://doi.org/http://dx.doi.org/10.1103/PhysRevLett.111.160601} {\bibfield
   {journal} {\bibinfo  {journal} {Phys. Rev. Lett.}\ }\textbf {\bibinfo
  {volume} {111}},\ \bibinfo {pages} {160601} (\bibinfo {year}
  {2013})}\BibitemShut {NoStop}%
\bibitem [{\citenamefont {Mallick}\ and\ \citenamefont
  {Chandrashekar}(2016)}]{mallick2016}%
  \BibitemOpen
  \bibfield  {author} {\bibinfo {author} {\bibfnamefont {A.}~\bibnamefont
  {Mallick}}\ and\ \bibinfo {author} {\bibfnamefont {C.~M.}\ \bibnamefont
  {Chandrashekar}},\ }\bibfield  {title} {\bibinfo {title} {Dirac cellular
  automaton from split-step quantum walk},\ }\href
  {https://doi.org/https://doi.org/10.1038/srep25779} {\bibfield  {journal}
  {\bibinfo  {journal} {Scientific reports}\ }\textbf {\bibinfo {volume} {6}},\
  \bibinfo {pages} {1} (\bibinfo {year} {2016})}\BibitemShut {NoStop}%
\bibitem [{\citenamefont {Devlin}\ \emph {et~al.}(2017)\citenamefont {Devlin},
  \citenamefont {Ambrosio}, \citenamefont {Rubin}, \citenamefont {Mueller},\
  and\ \citenamefont {Capasso}}]{devlin2017}%
  \BibitemOpen
  \bibfield  {author} {\bibinfo {author} {\bibfnamefont {R.~C.}\ \bibnamefont
  {Devlin}}, \bibinfo {author} {\bibfnamefont {A.}~\bibnamefont {Ambrosio}},
  \bibinfo {author} {\bibfnamefont {N.~A.}\ \bibnamefont {Rubin}}, \bibinfo
  {author} {\bibfnamefont {J.~P.~B.}\ \bibnamefont {Mueller}},\ and\ \bibinfo
  {author} {\bibfnamefont {F.}~\bibnamefont {Capasso}},\ }\bibfield  {title}
  {\bibinfo {title} {Arbitrary spin-to--orbital angular momentum conversion of
  light},\ }\href {https://doi.org/https://doi.org/10.1126/science.aao5392}
  {\bibfield  {journal} {\bibinfo  {journal} {Science}\ }\textbf {\bibinfo
  {volume} {358}},\ \bibinfo {pages} {896} (\bibinfo {year}
  {2017})}\BibitemShut {NoStop}%
\bibitem [{\citenamefont {Mueller}\ \emph {et~al.}(2017)\citenamefont
  {Mueller}, \citenamefont {Rubin}, \citenamefont {Devlin}, \citenamefont
  {Groever},\ and\ \citenamefont {Capasso}}]{mueller2017}%
  \BibitemOpen
  \bibfield  {author} {\bibinfo {author} {\bibfnamefont {J.~P.~B.}\
  \bibnamefont {Mueller}}, \bibinfo {author} {\bibfnamefont {N.~A.}\
  \bibnamefont {Rubin}}, \bibinfo {author} {\bibfnamefont {R.~C.}\ \bibnamefont
  {Devlin}}, \bibinfo {author} {\bibfnamefont {B.}~\bibnamefont {Groever}},\
  and\ \bibinfo {author} {\bibfnamefont {F.}~\bibnamefont {Capasso}},\
  }\bibfield  {title} {\bibinfo {title} {Metasurface polarization optics:
  independent phase control of arbitrary orthogonal states of polarization},\
  }\href {https://doi.org/http://dx.doi.org/10.1103/PhysRevLett.118.113901}
  {\bibfield  {journal} {\bibinfo  {journal} {Phys. Rev. Lett.}\ }\textbf
  {\bibinfo {volume} {118}},\ \bibinfo {pages} {113901} (\bibinfo {year}
  {2017})}\BibitemShut {NoStop}%
\bibitem [{\citenamefont {Allen}\ \emph {et~al.}(1992)\citenamefont {Allen},
  \citenamefont {Beijersbergen}, \citenamefont {Spreeuw},\ and\ \citenamefont
  {Woerdman}}]{allen92}%
  \BibitemOpen
  \bibfield  {author} {\bibinfo {author} {\bibfnamefont {L.}~\bibnamefont
  {Allen}}, \bibinfo {author} {\bibfnamefont {M.~W.}\ \bibnamefont
  {Beijersbergen}}, \bibinfo {author} {\bibfnamefont {R.~J.~C.}\ \bibnamefont
  {Spreeuw}},\ and\ \bibinfo {author} {\bibfnamefont {J.~P.}\ \bibnamefont
  {Woerdman}},\ }\bibfield  {title} {\bibinfo {title} {Orbital angular momentum
  of light and the transformation of laguerre-gaussian laser modes},\ }\href
  {https://doi.org/https://doi.org/10.1103/PhysRevA.45.8185} {\bibfield
  {journal} {\bibinfo  {journal} {Phys. Rev. A}\ }\textbf {\bibinfo {volume}
  {45}},\ \bibinfo {pages} {8185} (\bibinfo {year} {1992})}\BibitemShut
  {NoStop}%
\bibitem [{\citenamefont {Saleh}\ and\ \citenamefont
  {Teich}(2007)}]{saleh2007}%
  \BibitemOpen
  \bibfield  {author} {\bibinfo {author} {\bibfnamefont {B.~E.~A.}\
  \bibnamefont {Saleh}}\ and\ \bibinfo {author} {\bibfnamefont {M.~C.}\
  \bibnamefont {Teich}},\ }\href@noop {} {\emph {\bibinfo {title} {Fundamentals
  of photonics\,(II edition)}}}\ (\bibinfo  {publisher} {Wiley-Interscience},\
  \bibinfo {year} {2007})\BibitemShut {NoStop}%
\bibitem [{\citenamefont {Heckenberg}\ \emph {et~al.}(1992)\citenamefont
  {Heckenberg}, \citenamefont {McDuff}, \citenamefont {Smith},\ and\
  \citenamefont {White}}]{heckenberg92}%
  \BibitemOpen
  \bibfield  {author} {\bibinfo {author} {\bibfnamefont {N.~R.}\ \bibnamefont
  {Heckenberg}}, \bibinfo {author} {\bibfnamefont {R.}~\bibnamefont {McDuff}},
  \bibinfo {author} {\bibfnamefont {C.~P.}\ \bibnamefont {Smith}},\ and\
  \bibinfo {author} {\bibfnamefont {A.~G.}\ \bibnamefont {White}},\ }\bibfield
  {title} {\bibinfo {title} {Generation of optical phase singularities by
  computer-generated holograms},\ }\href
  {https://doi.org/https://doi.org/10.1364/OL.17.000221} {\bibfield  {journal}
  {\bibinfo  {journal} {Opt. Lett.}\ }\textbf {\bibinfo {volume} {17}},\
  \bibinfo {pages} {221} (\bibinfo {year} {1992})}\BibitemShut {NoStop}%
\bibitem [{\citenamefont {Berkhout}\ \emph {et~al.}(2010)\citenamefont
  {Berkhout}, \citenamefont {Lavery}, \citenamefont {Courtial}, \citenamefont
  {Beijersbergen},\ and\ \citenamefont {Padgett}}]{berkhout2010}%
  \BibitemOpen
  \bibfield  {author} {\bibinfo {author} {\bibfnamefont {G.~C.~G.}\
  \bibnamefont {Berkhout}}, \bibinfo {author} {\bibfnamefont {M.~P.~J.}\
  \bibnamefont {Lavery}}, \bibinfo {author} {\bibfnamefont {J.}~\bibnamefont
  {Courtial}}, \bibinfo {author} {\bibfnamefont {M.~W.}\ \bibnamefont
  {Beijersbergen}},\ and\ \bibinfo {author} {\bibfnamefont {M.~J.}\
  \bibnamefont {Padgett}},\ }\bibfield  {title} {\bibinfo {title} {Efficient
  sorting of orbital angular momentum states of light},\ }\href
  {https://doi.org/https://doi.org/10.1103/PhysRevLett.105.153601} {\bibfield
  {journal} {\bibinfo  {journal} {Phys. Rev. Lett.}\ }\textbf {\bibinfo
  {volume} {105}},\ \bibinfo {pages} {153601} (\bibinfo {year}
  {2010})}\BibitemShut {NoStop}%
\bibitem [{\citenamefont {Mirhosseini}\ \emph {et~al.}(2013)\citenamefont
  {Mirhosseini}, \citenamefont {Malik}, \citenamefont {Shi},\ and\
  \citenamefont {Boyd}}]{mirhosseini2013}%
  \BibitemOpen
  \bibfield  {author} {\bibinfo {author} {\bibfnamefont {M.}~\bibnamefont
  {Mirhosseini}}, \bibinfo {author} {\bibfnamefont {M.}~\bibnamefont {Malik}},
  \bibinfo {author} {\bibfnamefont {Z.}~\bibnamefont {Shi}},\ and\ \bibinfo
  {author} {\bibfnamefont {R.~W.}\ \bibnamefont {Boyd}},\ }\bibfield  {title}
  {\bibinfo {title} {Efficient separation of the orbital angular momentum
  eigenstates of light},\ }\href
  {https://doi.org/https://doi.org/10.1038/ncomms3781} {\bibfield  {journal}
  {\bibinfo  {journal} {Nat. Commun.}\ }\textbf {\bibinfo {volume} {4}},\
  \bibinfo {pages} {1} (\bibinfo {year} {2013})}\BibitemShut {NoStop}%
\bibitem [{\citenamefont {Barakat}(1993)}]{barakat93}%
  \BibitemOpen
  \bibfield  {author} {\bibinfo {author} {\bibfnamefont {R.}~\bibnamefont
  {Barakat}},\ }\bibfield  {title} {\bibinfo {title} {Analytic proofs of the
  arago--fresnel laws for the interference of polarized light},\ }\href
  {https://doi.org/https://doi.org/10.1364/JOSAA.10.000180} {\bibfield
  {journal} {\bibinfo  {journal} {J. Opt. Soc. Am. A}\ }\textbf {\bibinfo
  {volume} {10}},\ \bibinfo {pages} {180} (\bibinfo {year} {1993})}\BibitemShut
  {NoStop}%
\bibitem [{\citenamefont {Fickler}\ \emph {et~al.}(2014)\citenamefont
  {Fickler}, \citenamefont {Lapkiewicz}, \citenamefont {Huber}, \citenamefont
  {Lavery}, \citenamefont {Padgett},\ and\ \citenamefont
  {Zeilinger}}]{fickler2014}%
  \BibitemOpen
  \bibfield  {author} {\bibinfo {author} {\bibfnamefont {R.}~\bibnamefont
  {Fickler}}, \bibinfo {author} {\bibfnamefont {R.}~\bibnamefont {Lapkiewicz}},
  \bibinfo {author} {\bibfnamefont {M.}~\bibnamefont {Huber}}, \bibinfo
  {author} {\bibfnamefont {M.~P.}\ \bibnamefont {Lavery}}, \bibinfo {author}
  {\bibfnamefont {M.~J.}\ \bibnamefont {Padgett}},\ and\ \bibinfo {author}
  {\bibfnamefont {A.}~\bibnamefont {Zeilinger}},\ }\bibfield  {title} {\bibinfo
  {title} {Interface between path and orbital angular momentum entanglement for
  high-dimensional photonic quantum information},\ }\href
  {https://doi.org/https://doi.org/10.1038/ncomms5502} {\bibfield  {journal}
  {\bibinfo  {journal} {Nat. Commun.}\ }\textbf {\bibinfo {volume} {5}},\
  \bibinfo {pages} {1} (\bibinfo {year} {2014})}\BibitemShut {NoStop}%
\bibitem [{\citenamefont {Ono}\ \emph {et~al.}(2020)\citenamefont {Ono},
  \citenamefont {Hata}, \citenamefont {Tsunekawa}, \citenamefont {Nozaki},
  \citenamefont {Sumikura}, \citenamefont {Chiba},\ and\ \citenamefont
  {Notomi}}]{ono2020}%
  \BibitemOpen
  \bibfield  {author} {\bibinfo {author} {\bibfnamefont {M.}~\bibnamefont
  {Ono}}, \bibinfo {author} {\bibfnamefont {M.}~\bibnamefont {Hata}}, \bibinfo
  {author} {\bibfnamefont {M.}~\bibnamefont {Tsunekawa}}, \bibinfo {author}
  {\bibfnamefont {K.}~\bibnamefont {Nozaki}}, \bibinfo {author} {\bibfnamefont
  {H.}~\bibnamefont {Sumikura}}, \bibinfo {author} {\bibfnamefont
  {H.}~\bibnamefont {Chiba}},\ and\ \bibinfo {author} {\bibfnamefont
  {M.}~\bibnamefont {Notomi}},\ }\bibfield  {title} {\bibinfo {title}
  {Ultrafast and energy-efficient all-optical switching with graphene-loaded
  deep-subwavelength plasmonic waveguides},\ }\href
  {https://doi.org/https://doi.org/10.1038/s41566-019-0547-7} {\bibfield
  {journal} {\bibinfo  {journal} {Nat. Photonics}\ }\textbf {\bibinfo {volume}
  {14}},\ \bibinfo {pages} {37} (\bibinfo {year} {2020})}\BibitemShut {NoStop}%
\bibitem [{\citenamefont {Guo}\ \emph {et~al.}(2022)\citenamefont {Guo},
  \citenamefont {Sekine}, \citenamefont {Ledezma}, \citenamefont {Nehra},
  \citenamefont {Dean}, \citenamefont {Roy}, \citenamefont {Gray},
  \citenamefont {Jahani},\ and\ \citenamefont {Marandi}}]{guo2022}%
  \BibitemOpen
  \bibfield  {author} {\bibinfo {author} {\bibfnamefont {Q.}~\bibnamefont
  {Guo}}, \bibinfo {author} {\bibfnamefont {R.}~\bibnamefont {Sekine}},
  \bibinfo {author} {\bibfnamefont {L.}~\bibnamefont {Ledezma}}, \bibinfo
  {author} {\bibfnamefont {R.}~\bibnamefont {Nehra}}, \bibinfo {author}
  {\bibfnamefont {D.~J.}\ \bibnamefont {Dean}}, \bibinfo {author}
  {\bibfnamefont {A.}~\bibnamefont {Roy}}, \bibinfo {author} {\bibfnamefont
  {R.~M.}\ \bibnamefont {Gray}}, \bibinfo {author} {\bibfnamefont
  {S.}~\bibnamefont {Jahani}},\ and\ \bibinfo {author} {\bibfnamefont
  {A.}~\bibnamefont {Marandi}},\ }\bibfield  {title} {\bibinfo {title}
  {Femtojoule femtosecond all-optical switching in lithium niobate
  nanophotonics},\ }\href
  {https://doi.org/https://doi.org/10.1038/s41566-022-01044-5} {\bibfield
  {journal} {\bibinfo  {journal} {Nat. Photonics}\ }\textbf {\bibinfo {volume}
  {16}},\ \bibinfo {pages} {625} (\bibinfo {year} {2022})}\BibitemShut
  {NoStop}%
\bibitem [{\citenamefont {Singh}\ \emph
  {et~al.}(2021{\natexlab{b}})\citenamefont {Singh}, \citenamefont {Alderete},
  \citenamefont {Balu}, \citenamefont {Monroe}, \citenamefont {Linke},\ and\
  \citenamefont {Chandrashekar}}]{shivani2021}%
  \BibitemOpen
  \bibfield  {author} {\bibinfo {author} {\bibfnamefont {S.}~\bibnamefont
  {Singh}}, \bibinfo {author} {\bibfnamefont {C.~H.}\ \bibnamefont {Alderete}},
  \bibinfo {author} {\bibfnamefont {R.}~\bibnamefont {Balu}}, \bibinfo {author}
  {\bibfnamefont {C.}~\bibnamefont {Monroe}}, \bibinfo {author} {\bibfnamefont
  {N.~M.}\ \bibnamefont {Linke}},\ and\ \bibinfo {author} {\bibfnamefont
  {C.~M.}\ \bibnamefont {Chandrashekar}},\ }\bibfield  {title} {\bibinfo
  {title} {Quantum circuits for the realization of equivalent forms of
  one-dimensional discrete-time quantum walks on near-term quantum hardware},\
  }\href {https://doi.org/https://link.aps.org/doi/10.1103/PhysRevA.104.062401}
  {\bibfield  {journal} {\bibinfo  {journal} {Phys. Rev. A}\ }\textbf {\bibinfo
  {volume} {104}},\ \bibinfo {pages} {062401} (\bibinfo {year}
  {2021}{\natexlab{b}})}\BibitemShut {NoStop}%
\bibitem [{\citenamefont {Genske}\ \emph {et~al.}(2013)\citenamefont {Genske},
  \citenamefont {Alt}, \citenamefont {Steffen}, \citenamefont {Werner},
  \citenamefont {Werner}, \citenamefont {Meschede},\ and\ \citenamefont
  {Alberti}}]{MWA13}%
  \BibitemOpen
  \bibfield  {author} {\bibinfo {author} {\bibfnamefont {M.}~\bibnamefont
  {Genske}}, \bibinfo {author} {\bibfnamefont {W.}~\bibnamefont {Alt}},
  \bibinfo {author} {\bibfnamefont {A.}~\bibnamefont {Steffen}}, \bibinfo
  {author} {\bibfnamefont {A.~H.}\ \bibnamefont {Werner}}, \bibinfo {author}
  {\bibfnamefont {R.~F.}\ \bibnamefont {Werner}}, \bibinfo {author}
  {\bibfnamefont {D.}~\bibnamefont {Meschede}},\ and\ \bibinfo {author}
  {\bibfnamefont {A.}~\bibnamefont {Alberti}},\ }\bibfield  {title} {\bibinfo
  {title} {Electric quantum walks with individual atoms},\ }\href
  {https://doi.org/10.1103/PhysRevLett.110.190601} {\bibfield  {journal}
  {\bibinfo  {journal} {Phys. Rev. Lett.}\ }\textbf {\bibinfo {volume} {110}},\
  \bibinfo {pages} {190601} (\bibinfo {year} {2013})}\BibitemShut {NoStop}%
\bibitem [{\citenamefont {Hegde}\ and\ \citenamefont
  {Chandrashekar}(2022)}]{AC22}%
  \BibitemOpen
  \bibfield  {author} {\bibinfo {author} {\bibfnamefont {A.~S.}\ \bibnamefont
  {Hegde}}\ and\ \bibinfo {author} {\bibfnamefont {C.~M.}\ \bibnamefont
  {Chandrashekar}},\ }\bibfield  {title} {\bibinfo {title} {Characterization of
  anomalous diffusion in one-dimensional quantum walks},\ }\href
  {https://doi.org/10.1088/1751-8121/ac6b67} {\bibfield  {journal} {\bibinfo
  {journal} {Journal of Physics A: Mathematical and Theoretical}\ }\textbf
  {\bibinfo {volume} {55}},\ \bibinfo {pages} {234006} (\bibinfo {year}
  {2022})}\BibitemShut {NoStop}%
\bibitem [{\citenamefont {Wei}\ \emph {et~al.}(2020)\citenamefont {Wei},
  \citenamefont {Earl}, \citenamefont {Lin}, \citenamefont {Kou},\ and\
  \citenamefont {Yuan}}]{wei2020}%
  \BibitemOpen
  \bibfield  {author} {\bibinfo {author} {\bibfnamefont {S.}~\bibnamefont
  {Wei}}, \bibinfo {author} {\bibfnamefont {S.~K.}\ \bibnamefont {Earl}},
  \bibinfo {author} {\bibfnamefont {J.}~\bibnamefont {Lin}}, \bibinfo {author}
  {\bibfnamefont {S.~S.}\ \bibnamefont {Kou}},\ and\ \bibinfo {author}
  {\bibfnamefont {X.-C.}\ \bibnamefont {Yuan}},\ }\bibfield  {title} {\bibinfo
  {title} {Active sorting of orbital angular momentum states of light with a
  cascaded tunable resonator},\ }\href
  {https://doi.org/https://doi.org/10.1038/s41377-020-0243-x} {\bibfield
  {journal} {\bibinfo  {journal} {Light Sci. Appl.}\ }\textbf {\bibinfo
  {volume} {9}},\ \bibinfo {pages} {1} (\bibinfo {year} {2020})}\BibitemShut
  {NoStop}%
\bibitem [{\citenamefont {Ruffato}\ \emph {et~al.}(2018)\citenamefont
  {Ruffato}, \citenamefont {Girardi}, \citenamefont {Massari}, \citenamefont
  {Mafakheri}, \citenamefont {Sephton}, \citenamefont {Capaldo}, \citenamefont
  {Forbes},\ and\ \citenamefont {Romanato}}]{ruffato2018}%
  \BibitemOpen
  \bibfield  {author} {\bibinfo {author} {\bibfnamefont {G.}~\bibnamefont
  {Ruffato}}, \bibinfo {author} {\bibfnamefont {M.}~\bibnamefont {Girardi}},
  \bibinfo {author} {\bibfnamefont {M.}~\bibnamefont {Massari}}, \bibinfo
  {author} {\bibfnamefont {E.}~\bibnamefont {Mafakheri}}, \bibinfo {author}
  {\bibfnamefont {B.}~\bibnamefont {Sephton}}, \bibinfo {author} {\bibfnamefont
  {P.}~\bibnamefont {Capaldo}}, \bibinfo {author} {\bibfnamefont
  {A.}~\bibnamefont {Forbes}},\ and\ \bibinfo {author} {\bibfnamefont
  {F.}~\bibnamefont {Romanato}},\ }\bibfield  {title} {\bibinfo {title} {A
  compact diffractive sorter for high-resolution demultiplexing of orbital
  angular momentum beams},\ }\href
  {https://doi.org/https://doi.org/10.1038/s41598-018-28447-1} {\bibfield
  {journal} {\bibinfo  {journal} {Sci. Rep.}\ }\textbf {\bibinfo {volume}
  {8}},\ \bibinfo {pages} {1} (\bibinfo {year} {2018})}\BibitemShut {NoStop}%
\bibitem [{\citenamefont {Li}\ and\ \citenamefont {Zhao}(2017)}]{li2017}%
  \BibitemOpen
  \bibfield  {author} {\bibinfo {author} {\bibfnamefont {C.}~\bibnamefont
  {Li}}\ and\ \bibinfo {author} {\bibfnamefont {S.}~\bibnamefont {Zhao}},\
  }\bibfield  {title} {\bibinfo {title} {Efficient separating orbital angular
  momentum mode with radial varying phase},\ }\href
  {https://doi.org/https://doi.org/10.1364/PRJ.5.000267} {\bibfield  {journal}
  {\bibinfo  {journal} {Photonics Res.}\ }\textbf {\bibinfo {volume} {5}},\
  \bibinfo {pages} {267} (\bibinfo {year} {2017})}\BibitemShut {NoStop}%
\bibitem [{\citenamefont {Li}\ \emph {et~al.}(2019)\citenamefont {Li},
  \citenamefont {Xu}, \citenamefont {Maruthiyodan~Veetil}, \citenamefont
  {Valuckas}, \citenamefont {Paniagua-Dom{\'\i}nguez},\ and\ \citenamefont
  {Kuznetsov}}]{li2019}%
  \BibitemOpen
  \bibfield  {author} {\bibinfo {author} {\bibfnamefont {S.-Q.}\ \bibnamefont
  {Li}}, \bibinfo {author} {\bibfnamefont {X.}~\bibnamefont {Xu}}, \bibinfo
  {author} {\bibfnamefont {R.}~\bibnamefont {Maruthiyodan~Veetil}}, \bibinfo
  {author} {\bibfnamefont {V.}~\bibnamefont {Valuckas}}, \bibinfo {author}
  {\bibfnamefont {R.}~\bibnamefont {Paniagua-Dom{\'\i}nguez}},\ and\ \bibinfo
  {author} {\bibfnamefont {A.~I.}\ \bibnamefont {Kuznetsov}},\ }\bibfield
  {title} {\bibinfo {title} {Phase-only transmissive spatial light modulator
  based on tunable dielectric metasurface},\ }\href
  {https://doi.org/https://doi.org/10.1126/science.aaw6747} {\bibfield
  {journal} {\bibinfo  {journal} {Science}\ }\textbf {\bibinfo {volume}
  {364}},\ \bibinfo {pages} {1087} (\bibinfo {year} {2019})}\BibitemShut
  {NoStop}%
\bibitem [{\citenamefont {Roy}\ \emph {et~al.}(2018)\citenamefont {Roy},
  \citenamefont {Zhang}, \citenamefont {Jung}, \citenamefont {Troccoli},
  \citenamefont {Capasso},\ and\ \citenamefont {Lopez}}]{roy2018}%
  \BibitemOpen
  \bibfield  {author} {\bibinfo {author} {\bibfnamefont {T.}~\bibnamefont
  {Roy}}, \bibinfo {author} {\bibfnamefont {S.}~\bibnamefont {Zhang}}, \bibinfo
  {author} {\bibfnamefont {I.~W.}\ \bibnamefont {Jung}}, \bibinfo {author}
  {\bibfnamefont {M.}~\bibnamefont {Troccoli}}, \bibinfo {author}
  {\bibfnamefont {F.}~\bibnamefont {Capasso}},\ and\ \bibinfo {author}
  {\bibfnamefont {D.}~\bibnamefont {Lopez}},\ }\bibfield  {title} {\bibinfo
  {title} {Dynamic metasurface lens based on mems technology},\ }\href
  {https://doi.org/https://doi.org/10.1063/1.5018865} {\bibfield  {journal}
  {\bibinfo  {journal} {APL Photonics}\ }\textbf {\bibinfo {volume} {3}},\
  \bibinfo {pages} {021302} (\bibinfo {year} {2018})}\BibitemShut {NoStop}%
\bibitem [{\citenamefont {Shirmanesh}\ \emph {et~al.}(2020)\citenamefont
  {Shirmanesh}, \citenamefont {Sokhoyan}, \citenamefont {Wu},\ and\
  \citenamefont {Atwater}}]{shirmanesh2020}%
  \BibitemOpen
  \bibfield  {author} {\bibinfo {author} {\bibfnamefont {G.~K.}\ \bibnamefont
  {Shirmanesh}}, \bibinfo {author} {\bibfnamefont {R.}~\bibnamefont
  {Sokhoyan}}, \bibinfo {author} {\bibfnamefont {P.~C.}\ \bibnamefont {Wu}},\
  and\ \bibinfo {author} {\bibfnamefont {H.~A.}\ \bibnamefont {Atwater}},\
  }\bibfield  {title} {\bibinfo {title} {Electro-optically tunable
  multifunctional metasurfaces},\ }\href
  {https://doi.org/https://doi.org/10.1021/acsnano.0c01269} {\bibfield
  {journal} {\bibinfo  {journal} {ACS Nano}\ }\textbf {\bibinfo {volume}
  {14}},\ \bibinfo {pages} {6912} (\bibinfo {year} {2020})},\ \bibinfo {note}
  {pMID: 32352740}\BibitemShut {NoStop}%
\end{thebibliography}

%

\end{document}